\documentclass[twocolumn,aps,prb,superscriptaddress]{revtex4-2}

\usepackage[utf8]{inputenc}
\usepackage[version=4]{mhchem}
\usepackage{textcomp}
\usepackage{graphicx}
\usepackage{float}
\usepackage[caption=false]{subfig}
\usepackage{tabularx}
\usepackage{booktabs}
\usepackage{textcomp}
\usepackage{multirow}
\usepackage{placeins}

\newcommand{\degC}{\textdegree{}C}
\newcommand{\degph}{\textdegree{}C/h}

\begin{document}

\title{Polymorphic \ce{PtBi2}: Growth, structure and superconducting properties}

\author{G.~Shipunov}
\email[]{g.shipunov@ifw-dresden.de}
\affiliation{Leibniz Institute for Solid State and Materials Research Dresden, Helmholtzstr.\ 20, D-01069 Dresden, Germany}
\author{I.~Kovalchuk}
\affiliation{Leibniz Institute for Solid State and Materials Research Dresden, Helmholtzstr.\ 20, D-01069 Dresden, Germany}
\affiliation{Kyiv Academic University, 03142 Kyiv, Ukraine}
\author{B.~R.~Piening}
\author{V.~Labracherie}
\author{A.~Veyrat}
\author{D.~Wolf}
\author{A.~Lubk}
\author{S.~Subakti}
\affiliation{Leibniz Institute for Solid State and Materials Research Dresden, Helmholtzstr.\ 20, D-01069 Dresden, Germany}
\author{R.~Giraud}
\affiliation{Leibniz Institute for Solid State and Materials Research Dresden, Helmholtzstr.\ 20, D-01069 Dresden, Germany}
\affiliation{Université Grenoble Alpes, CNRS, CEA, Spintec, F-38000 Grenoble, France}
\author{J.~Dufouleur}
\author{S.~Shokri}
\author{F.~Caglieris}
\author{C.~Hess}
\author{D.~V.~Efremov}
\affiliation{Leibniz Institute for Solid State and Materials Research Dresden, Helmholtzstr.\ 20, D-01069 Dresden, Germany}
\author{B.~B\"uchner}
\affiliation{Leibniz Institute for Solid State and Materials Research Dresden, Helmholtzstr.\ 20, D-01069 Dresden, Germany}
\affiliation{Institut f\"ur Festk\"orperphysik, TU Dresden, D-01062 Dresden, Germany}
\author{S.~Aswartham}
\email[]{s.aswartham@ifw-dresden.de}
\affiliation{Leibniz Institute for Solid State and Materials Research Dresden, Helmholtzstr.\ 20, D-01069 Dresden, Germany}

\date{\today}

\begin{abstract}
  \ce{PtBi2} is a polymorphic system with interesting electronic properties.
  Here we report optimized crystal growth and structural characterization of pyrite-type and trigonal modification of \ce{PtBi2}.
  Selected area electron diffraction, X-ray powder diffraction and further Rietveld refinement confirms that trigonal \ce{PtBi2} crystallizes in non-centrosymmetric $P31m$ space group, pyrite-type \ce{PtBi2} in $Pa\bar{3}$ space group.
  Series of \ce{Pt_{1-x}Rh_xBi2} samples was obtained for $x=0, 0.03, 0.35$ in the trigonal \ce{PtBi2} structure.
  These \ce{Pt_{1-x}Rh_xBi2} compounds become superconducting where
  critical temperature increases from $T_c=600$\,mK  for $x=0$ up to $T_c=2.7$\,K for $x=0.35$.
  Furthermore we calculate the electronic band structure, using the structure parameters obtained.
  The calculated density of states (DOS) shows a minimum for the stochiometric compound at the Fermi level.
  These findings warrant further research by broader array of experimental techniques, as well as the effect of the substitution on the non-trivial band structure.
\end{abstract}


\maketitle

\section{Introduction}

Topological materials (TM) are a new class of quantum materials, which are characterized by a non-trivial topological band structure~\cite{lee2019topological}.
After initial discovery of such properties in the family of topological insulators, many other types of TMs followed, including Dirac and Weyl types of topological semimetals (TSM), which are characterized by conduction and valence band touching at several points near the Fermi level and show linear electron dispersion near those points, which are termed Dirac and Weyl nodes respectively~\cite{savage2018topology,yan2017topological}.
The presence of these points was experimentally detected by ARPES measurements as Fermi arcs~\cite{xu2015discovery}.
These topological bands strongly influence charge transport properties, such as electron mobility, giant magnetoresistance and anomalous Hall effect~\cite{borisenko2014experimental, shekhar2015extremely, liang2015ultrahigh,arnold2016negative,zyuzin2016intrinsic}.
Combination of non-trivial band structure with superconductivity in the same system makes it even more interesting due to possibility of realization of Majorana fermions~\cite{vaezi2013fractional}.

In recent years \ce{PtBi2} attracted a lot of attention from scientific community as one of the members of TSM family, which together with report of superconductivity in the system~\cite{alekseevski1953} makes it an attractive candidate for topological superconductivity.
It crystallizes in 4 polymorphic modifications~\cite{okamoto1991bi}: $\delta$, that could be formed by peritectic reaction at 660\degC{}, which is thermodynamically stable down to 640\degC{}. Temperature range of 420--640\degC{} corresponds to the $\gamma$-modification~\cite{zhu1, okamoto1991bi}, in between 270\degC{} and 420\degC{} $\beta$-modification is thermodynamically favorable with final polymorphic transition into the $\alpha$-modification at the temperature around 270\degC{}~\cite{okamoto1991bi, bar}.

\begin{figure}
  \centering{}
  \includegraphics[width=\columnwidth{}]{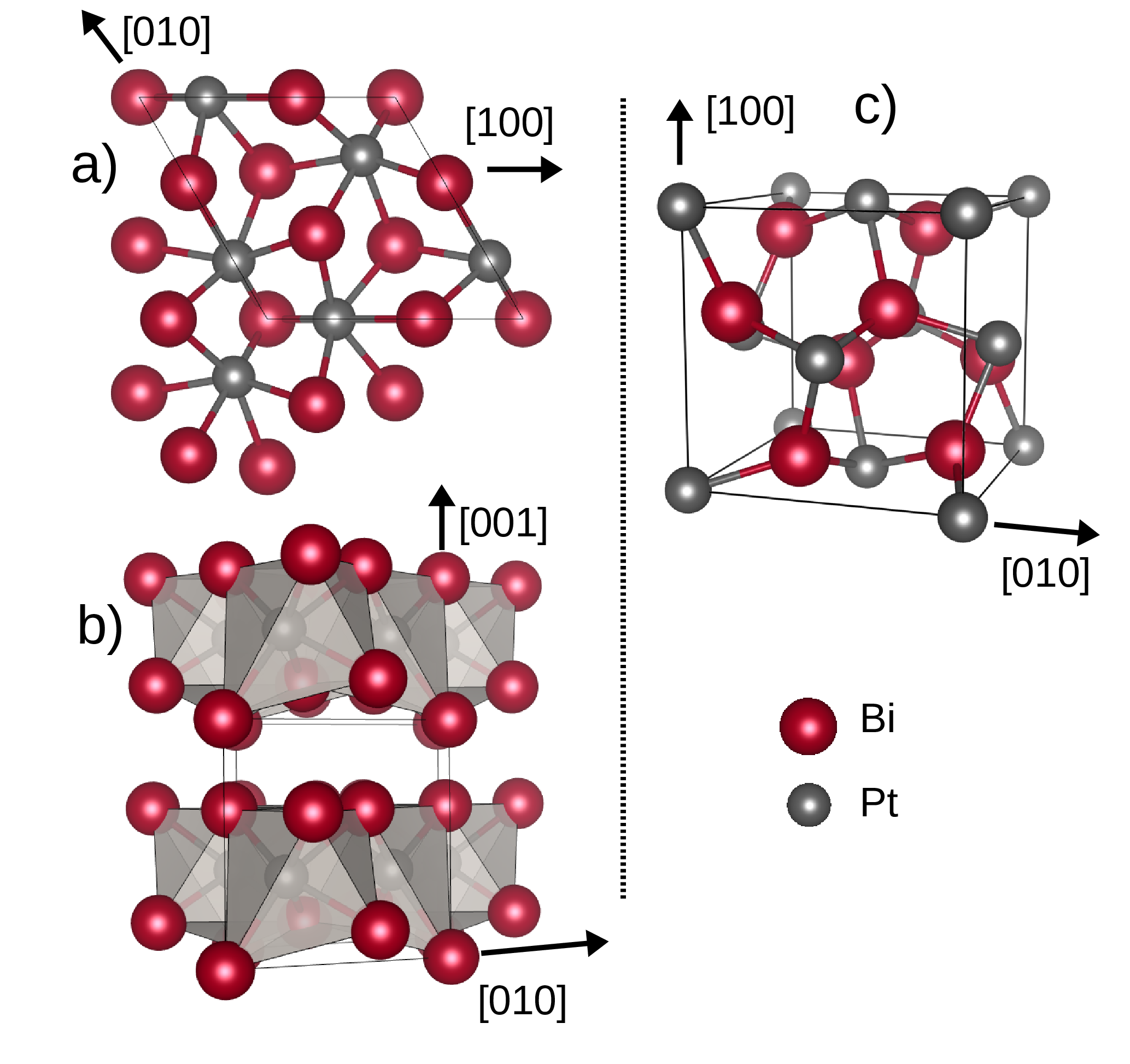}
  \caption{\label{structure}
    Reported structures of trigonal (a, b) and pyrite-type (c) polymorphs of \ce{PtBi2}.
    (a) Trigonal polymorph, model in space group $P31m$ proposed in~\cite{kaiser}, view along [001];
    (b) Same, view along [100];
    (c) Reported structure of pyrite-type polymorph of \ce{PtBi2} (space group $Pa\bar{3}$)~\cite{brese1994bonding}.
}
\end{figure}

Two of these modifications, $\beta$ (cubic, usually referred as pyrite-type) and $\gamma$ (hexagonal, referred as trigonal below) were recently shown to exhibit interesting physical properties.
Pyrite-type \ce{PtBi2} (space group $Pa\bar{3}$) shows extremely large unsaturated magnetoresistance, superseding the values demonstrated by \ce{WTe2}~\cite{gao2017extremely}, have been proposed to host Dirac fermions~\cite{PhysRevB.91.205128}.
Further multiband superconductivity with perfect electron-hole compensation under high pressure was reported~\cite{PhysRevMaterials.2.054203}.
Pyrite-type \ce{PtBi2} was reported to host a sixfold fermion near the Fermi level, with confirmation from ARPES~\cite{2020arXiv200608642T}.
Trigonal polymorph is an example of layred van der Waals material, namely 2D layers of covalently bound \ce{Pt} and \ce{Bi} are held together weakly, which makes this system a promising candidate for thickness depended studies due to ease of exfoliation.
This modification also shows large magnetoresistance~\cite{triclinMR}, however, no superconductivity was reported for this modification.
There are different reports on theoretical band structure of trigonal \ce{PtBi2}: some reports assume centrosymmetric space group $P\bar{3}$~\cite{thirupathaiah2018,yao2016,xu2016}, while in the ref.~\cite{yao2016} non-centrosymmetric space group $P31m$ was assumed.
Both structure variations were reported previously with the same lattice parameters:
$a=6.57$\,\AA{}, $c=6.16$\,\AA{} for $P\bar{3}$~\cite{bar} (ICSD \textnumero{}58847) and $a=6.573$\,\AA{}, $c=6.167$\,\AA{} for $P31m$~\cite{kaiser} (ICSD \textnumero{}428088), images of reported structures are presented on fig.~\ref{structure} for clarity.
In $P\bar{3}$ model \ce{Pt} atoms are laid out in a corrugated triangular lattice, with every \ce{Pt} atom coordinated by 6 atoms of \ce{Bi}, while in $P31m$ model \ce{Pt} position shifted towards the $3$ axis to complete the \ce{Pt} outer shell to magic number $18e^-$, which results in further deformation of \ce{PtBi6} octahedra and corrugation of Bi layers.

Trigonal modification of \ce{PtBi2} was predicted to host Dirac fermions while described in space group $P\bar{3}$~\cite{thirupathaiah2018} or Weyl fermions and triply-degenerate point while described in $P31m$ space group due to absence of spatial inversion symmetry~\cite{Gao2018}, so distinction between the two is crucial for further investigation of physical properties of the material.
This, together with the fact that predicted triply-degenerate points are near the Fermi level~\cite{Gao2018}, makes trigonal \ce{PtBi2} an interesting material for probing the properties of such fermionic excitations.

The aim of the present article is to report optimized synthesis conditions, to study the crystal structure and physical properties of pristine \ce{PtBi2} as well as \ce{PtBi2} with substitution of Pt by Rh.
Substitution of Pt (with outer shell configuration $5d^9$) by Rh ($4d^8$) produces overall hole doping which should affect DOS near Fermi level.
The article is constructed as follows: in the first section we report optimized methods of crystal growth of both pyrite-type and trigonal modification. For trigonal modification we study a series of crystals \ce{Pt_{1-x}Rh_{x}Bi2} by means of SEM-EDX, XRD, SAED, SQUID and electrical transport measurements.
Using obtained parameters of the crystal structure we find the electronic band structure for the trigonal polymorph in the framework of the DFT theory.
Also we show the superconductivity in trigonal polymorph with the way of enhancing transition temperature by Rh substitution.

\section{Methods}
\subsection{Crystal growth}
\begin{figure}
 \includegraphics[width=\columnwidth]{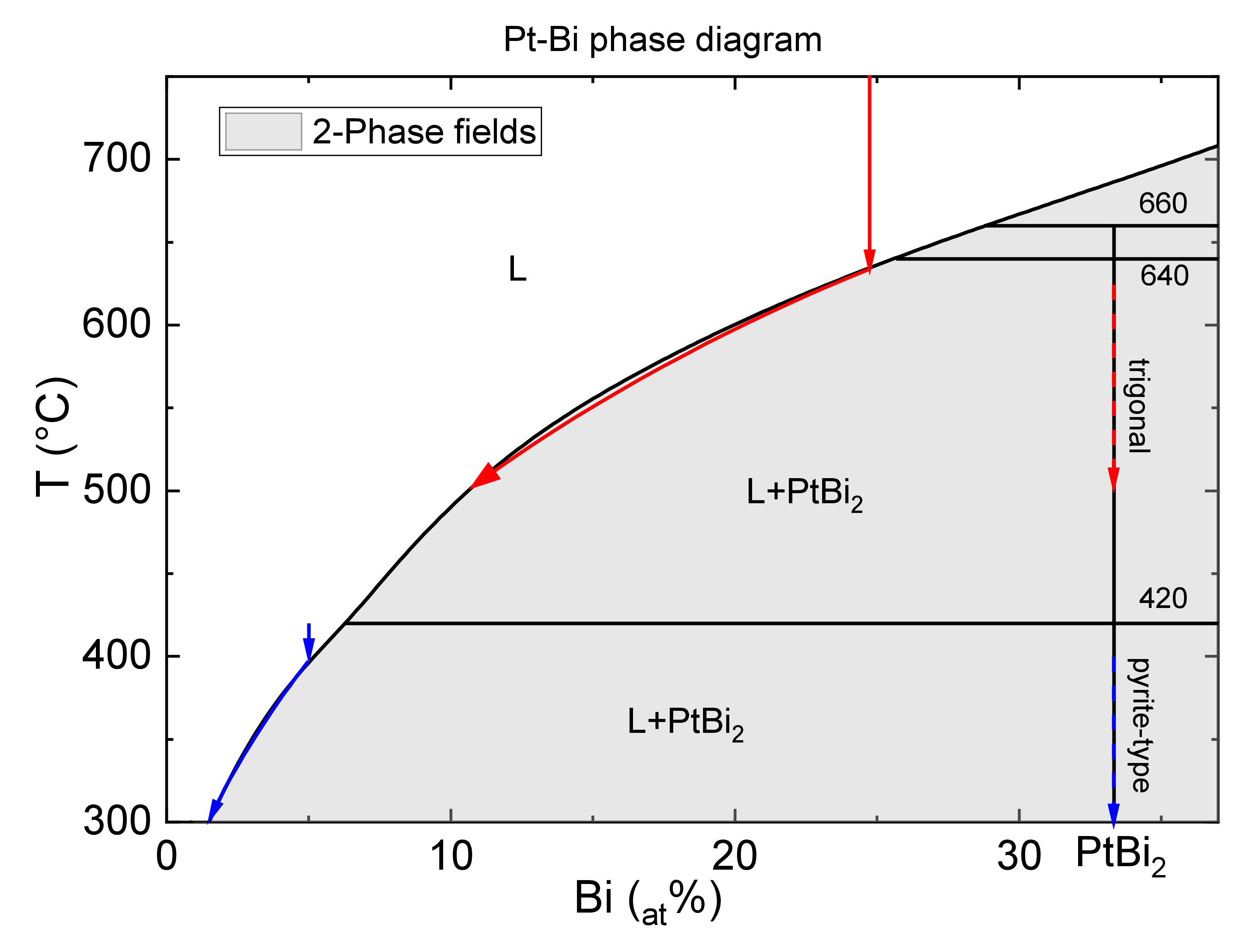}
 \caption{\label{img:pd}
 The fragment of published Pt-Bi phase diagram~\cite{okamoto1991bi}.
 The arrows schematically represent the synythesis conditions for trigonal~(red) and pyrite-type modification~(blue).
 Solid lines show composition of liquid phase during the synthesis, while dashed --- composition of percipitating solid phase.}
 \end{figure}

Single crystals of pyrite-type and trigonal \ce{PtBi2} were grown via self-flux method.
All ampules, crucibles and quartz wool were heat treated before the synthesis at 800\degC{} for 24\,hours.
The optimized temperatures were chosen according to the published Pt-Bi phase diagram~\cite{okamoto1991bi} (region of the phase diagram is presented in fig.~\ref{img:pd}), such that crystallization happens only in the crystallization zone of chosen polymorphic modification, preventing precipitation of unwanted modifications. According to the phase diagram it is possible to selectively crystallize trigonal \ce{PtBi2} from \ce{Bi} flux in the temperature region between 420--640\degC{} with corresponding \ce{Pt} content of 7--26\,at.\,\% (red lines on fig.~\ref{img:pd}),
and for pyrite-type modification in temperature region of 272--420\degC{} and \ce{Pt} content under 7\,at.\,\% (blue lines on fig.~\ref{img:pd}).
For \ce{Pt_{1-x}Rh_{x}Bi2} samples due to absence of published ternary phase diagram the similar temperature ranges and flux to metal ratios were assumed.
Small drops of flux residue were removed from the surface either mechanically or by etching in \ce{HNO3} dilute solution.

\paragraph*{Trigonal modification:} Single crystals of trigonal \ce{PtBi2} and \ce{Pt_{1-x}Rh_{x}Bi2} were synthesized by
mixing elemental powders of \ce{Pt} (99.99\%, Saxonia Edelmetalle GmbH), \ce{Rh} (99.95\%, Evochem) and \ce{Bi} (325-mesh powder, 99.5\%, Alfa Aesar).
elemental powders with molar ratio of \ce{(Pt_{1-x}Rh_{x})}:\ce{Bi}=1:4 (for $x=0$, $0.1$ and $0.3$) and a total charge mass of 0.4\,g were homogenized by manual grinding for 5\,min.\ in agate mortar and placed into a Canfield crucible set~\cite{canfield} to facilitate flux removal on later stage.
The crucible in turn was sealed inside of an evacuated quartz glass tube to prevent oxidation.
For the unsubstituted compound the setup then was heated to 850\degC{} at 100\degph{} and held at that temperature for 10\,h.
After that the mixture was cooled to 420\degC{} with a rate of 2\degph{}, after which excess of the flux was removed by centrifugation.
Rh-containing reaction mixtures were also heated to 850\degC{} at 100\degph{} with the same dwell time at maximum temperature of 10\,h,
but due to absence of phase diagram data in trenary (Pt--Rh--Bi) system the cooling step was shortened not to go below the temperature of 500\degC{}, to prevent crystallization of hypothetical secondary phases.
As grown crystals shown in fig.~\ref{photo}b and c, and clearly demonstrate the layered morphology.

\paragraph*{Pyrite-type modification:} Crystals of \ce{PtBi2} in pyrite-type modification were obtained in likewise manner:
Mixture of elemental powders with molar ratio of \ce{Pt}:\ce{Bi} $=1:20$ was homogenized by manual grinding in agate mortar, placed into Canfield crucible set and sealed in quartz ampule.
The setup was heated to 400\degC{} at 100\degph{}, held at that temperature for 4\,days and then cooled at 0.25\degph{} to 300\degC{}, with centrifugation of excess flux hereafter.
As grown crystals are faceted and isometric, as shown in fig.~\ref{photo}a.

\subsection{Characterization of composition and structure}
The composition of the as grown single crystals was determined by energy-dispersive X-ray spectroscopy (EDX),
with electron beam probe (accelerating voltage 30kV, current 552pA).
Scanning electron microscopy images are shown in fig.~\ref{photo}d--f.
Structural characterization and phase purity was confirmed by means of powder X-ray diffraction using STOE powder diffractometer ($2\theta$:$\omega$ scan, \ce{Co}~$K_{\alpha{}1}$ or \ce{Mo}~$K_{\alpha{}1}$ radiation, curved Ge (111) monochromator, Debye-Scherrer geometry).
Rietveld refinement of the x-ray data was carried out with FullProf~\cite{fullprof} and Jana2006~\cite{jana} software packages.
Selected area electron diffraction (SAED) on \ce{PtBi2} and \ce{Pt_{1-x}Rh_{x}Bi_2} nanoflakes was performed on an FEI Tecnai G2 transmission electron microscope (ThermoFisher Comp., US) with \ce{LaB6} emitter operated at 200\,kV acceleration voltage.
The quality and stability of \ce{Pt_{1-x}Rh_{x}Bi2} crystals used in this experiment allow us to carry out mechanically exfoliation process using commercially available adhesive tape (vivess).
The tape and exfoliated \ce{Pt_{1-x}Rh_{x}Bi2} crystals were separated by immersion in 20\,mL of acetone and isopropanol (1:1) solution in 50\,mL beaker glass.
Ultrasonic waves (frequency ~35\,KHz) was employed to assist the separation of \ce{Pt_{1-x}Rh_{x}Bi2}-flakes from the tape surface.
The micrometer size \ce{Pt_{1-x}Rh_{x}Bi2}-flakes for TEM characterizations were then transferred from the bottom of the beaker glass to the \ce{Cu}-grids using a standard pipette.
To ensure the cleanliness of the \ce{Pt_{1-x}Rh_{x}Bi2}-flakes for selected area diffraction investigation, we determined the elemental compositions of every \ce{Pt_{1-x}Rh_{x}Bi2}-flake by in-situ EDS before collecting the selected area diffraction pattern data.
Theoretical kinematic electron diffraction patterns were computed and visualized using the SingleCrystal software package version 3.1.5 (CrystalMaker Software LtD., UK).

\subsection{Characterization of physical properties}
In-plane resistivity measurements have been performed in a standard 4-probe configuration.
Electrical contacts have been made with copper or silver wires glued to the sample using a conducting silver paint (Dupont 4929n).
The measurements have been performed in the temperature range 2.3--300\,K in a liquid \ce{^4He} cryostat endowed with a 15\,T magnet and in temperature range 0.1--1\,K using a dilation fridge in an liquid \ce{^4He} cryostat with a 3D vector magnet (6T--2T--2T).

Magnetization data were measured using a Quantum Design MPMS SQUID with vibrating sample magnetometer.
The electronic band-structure was obtained in the framework of fully relativistic density functional theory (DFT) using the Full Potential Local Orbital band structure package (FPLO)~\cite{kopernik}.
The calculations were carried out within the generalized gradient approximation (GGA) of the Perdew-Burke-Ernzerhof (PBE) exchange-correlation potential~\cite{PhysRevLett.77.3865}.
A $k$-mesh of $12\times12\times12$ $k$-points in the whole Brillouin zone was employed.

\section{Results and Discussion}

\subsection{Composition and structure}
\begin{figure}
  \centering
  \includegraphics[width=\columnwidth{}]{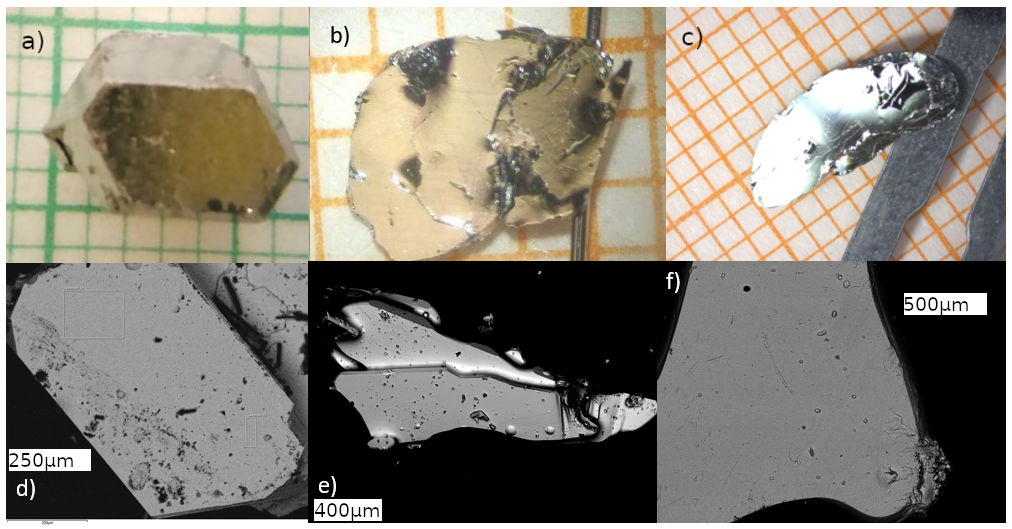}
  \caption{\label{photo}As grown crystals on millimeter scale: (a) pyrite-type \ce{PtBi2}; (b) trigonal \ce{PtBi2}; (c) \ce{Pt_{0.65}Rh_{0.35}Bi2}. SEM-BSE images: (d) \ce{PtBi2}; (e,f) \ce{Pt_{0.65}Rh_{0.35}Bi2}}
\end{figure}

Pyrite-type \ce{PtBi2} was obtained as large, well-faceted, isometric silvery crystals up to 0.5\,cm in diameter and a mass of up to 500\,mg.
As an example, one of the as-grown pyrite-type crystals is shown in fig.~\ref{photo}a.
Crystals of trigonal \ce{PtBi2} and \ce{Pt_{1-x}Rh_{x}Bi2} were obtained as easily cleavable silvery plates with a layered morphology in tabular hexagonal habit,
which is in line with layered van~der~Waals structure of the material.
The crystals of size of up to 5x10\,mm and a mass of up to 200\,mg were acquired as shown in fig.~\ref{photo}b and c.

\begin{table}
  \caption{\label{edx-table}
    Nominal compositions and compositions according to SEM-EDX for obtained \ce{Pt_{1-x}Rh_xBi2} compounds}
  \begin{tabular}{ll}
    \toprule{}
    nominal & SEM-EDX \\
    \midrule{}
    \emph{cubic:} &\\
    \ce{PtBi2} & \ce{PtBi_{2.00(2)}}\\
    \midrule{}
    \emph{trigonal:}&\\
    \ce{PtBi2} & \ce{PtBi_{2.03(4)}} \\
    \ce{Pt_{0.9}Rh_{0.1}Bi2} & \ce{Pt_{0.97(1)}Rh_{0.02(1)}Bi_{1.89(2)}} \\
    \ce{Pt_{0.7}Rh_{0.3}Bi2} & \ce{Pt_{0.64(2)}Rh_{0.35(1)}Bi_{1.9(3)}} \\
    \bottomrule{}
  \end{tabular}
\end{table}

\begin{figure*}
  \begin{centering}
    \includegraphics[width=\textwidth{}]{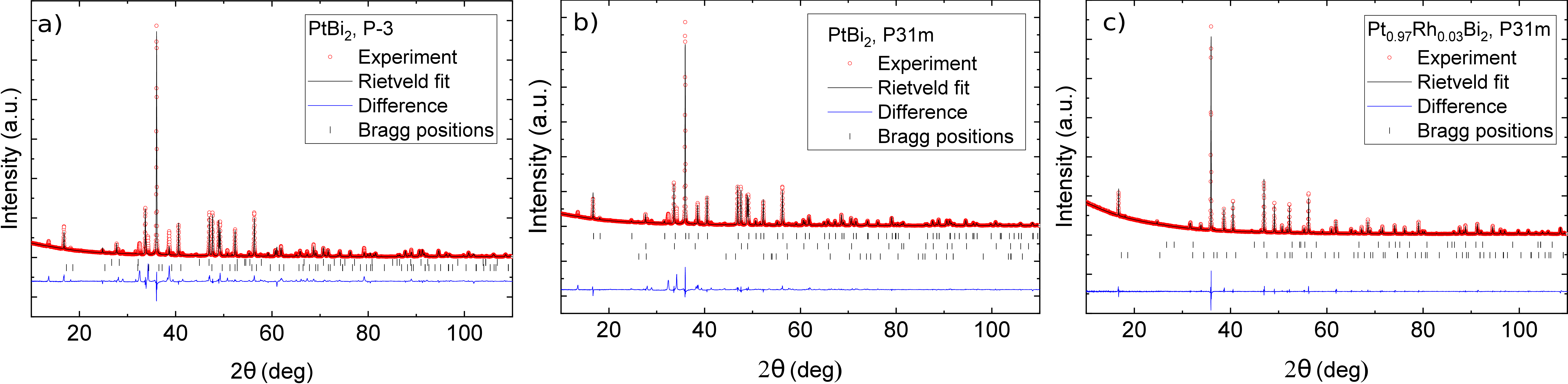}
    \caption{\label{riet}Rietveld analysis of \ce{Pt_{1-x}Rh_xBi2}. For $x=0$: (a) using model in $P\bar{3}$ space group~\cite{bar},
      (b) --- $P31m$ space group~\cite{kaiser}.
      (c) Rietveld fit for $x=0.03$ with model in $P31m$ space group.}
  \end{centering}
\end{figure*}

SEM-EDX analysis of both trigonal (for SEM images see fig.~\ref{photo}d and e) and pyrite-type modification of \ce{PtBi2} confirmed the stochiometric composition of the compounds and showed homogenous distribution of the elements.
Samples with substitution show uniform rhodium incorporation into the crystal.
For 10\% and 30\% nominal substitution level measured EDX composition is \ce{Pt_{0.97(1)}Rh_{0.02(1)}Bi_{1.89(2)}}
and \ce{Pt_{0.64(2)}Rh_{0.35(1)}Bi_{1.9(3)}} respectively.
Results of EDX analysis are presented in the table~\ref{edx-table}.
The structure of pyrite-type \ce{PtBi2} was determined by powder X-ray diffraction with subsequent Rietveld analysis.
X-ray powder analysis shows no secondary phases and obtained structural parameters are agreeing well with the ones published previously with $a=6.702$ in space group $Pa\bar{3}$~\cite{brese1994bonding}.

\begin{figure}
    \includegraphics[width=\columnwidth{}]{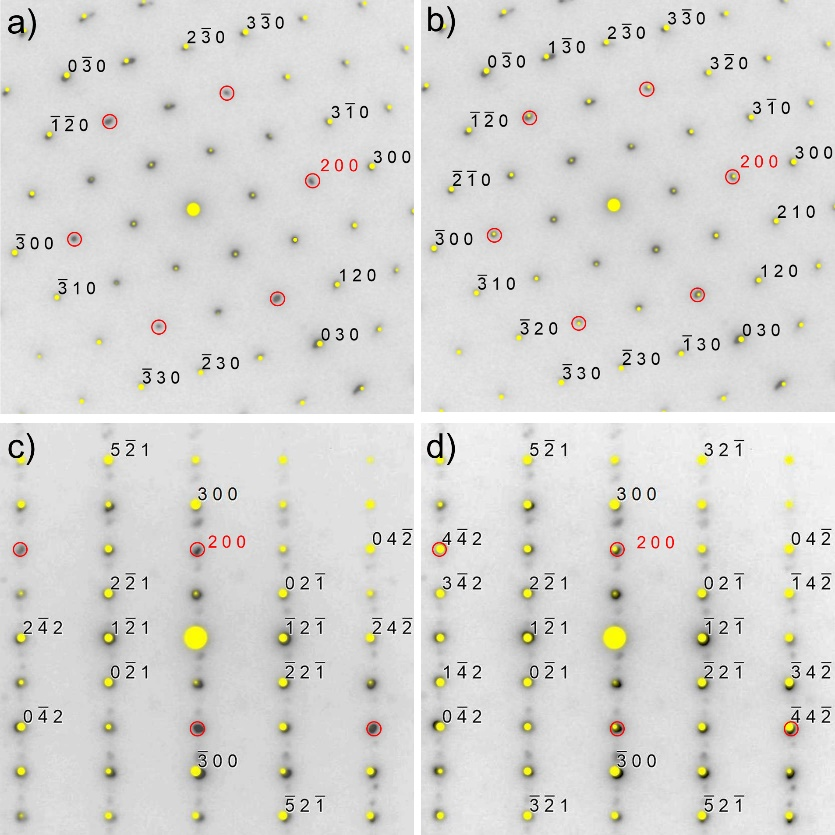}
    \caption{\label{TEM}
    Selected area electron diffraction (SAED) on \ce{Pt_{1-x}Rh_{x}Bi2}. (a,b) SAED pattern of a representative nanoflake (x=0) oriented in [001] zone axis overlaid with theoretical SAED patterns (yellow dots) using $P\bar{3}$ (a) and $P31m$ (b) space group structure. (c,d) SAED pattern of a representative nanoflake (x=0.35) oriented in [012] zone axis overlaid with theoretical SAED patterns (yellow dots) using $P\bar{3}$ (c) and $P31m$ (d) space group structure. The red circles in (a-d) indicate the 200 and symmetrical equivalent reflections that are virtually forbidden for $P\bar{3}$ symmetry (a,c), but clearly show up for $P31m$(b,d).}
\end{figure}

For trigonal modification powder XRD data, from crystals ground by hand show abnormally broad diffraction peaks.
This behaviour might be linked to the high ductility and ease of cleavage of the material.
To obtain high quality XRD data, crystals were ground in a ball mill for 30~min, and afterwards the powder was annealed at the centrifugation temperature to relieve any internal stress caused by milling, and quenched in ambient temperature water to prevent polymorphic transformation to pyrite-type modification.
Since two slightly different crystal structures for the trigonal modification were reported in the literature with same lattice parameters, our pattern was compared to the theoretically modeled one for \ce{PtBi2} structures reported in ICSD~\cite{bar,kaiser}.
The pattern shows minor quantity of secondary phases present.
Those secondary phases were fitted with le-Bail method prior the Rietveld refinement to exclude them from consideration.
Results of the Rietveld fit for trigonal \ce{PtBi2} are presented on fig.~\ref{riet}~(a) and~(b).
We performed the analysis of the X-Ray data employing both models presented in the literature: in $P\bar{3}$~\cite{bar} and $P31m$~\cite{kaiser}.
While in both cases the lattice parameters agree with the data published previously, refinement in $P\bar{3}$ space group gives rise to negative isotropic displacement parameters for ``Pt1'' and ``Bi1'' positions.
Moreover the residual parameters for this model are considerably worse for the $P\bar{3}$ model (e.g. $R=0.1163$ vs $R=0.0520$ for $P31m$ model).
Comparison of refinement results are presented in table~\ref{refine-table}.
To further support this finding selected area electron diffraction (SAED) on individual \ce{Pt_{1-x}Rh_{x}Bi2} flakes exfoliated from a single crystal was carried out (fig.~\ref{TEM}).
The SAED patterns for \ce{Pt_{1-x}Rh_{x}Bi2} recorded in [001] (x=0) and [012] (x=0.35) orientation were compared with theoretical patterns for both space groups $P\bar{3}$ fig.~\ref{TEM}(a,c) and $P31m$ fig.~\ref{TEM}(b,d). The presence of the 200 and symmetrical equivalent reflections in the experimental data strongly suggest a $P31m$ space group structure for both crystals. These reflections are forbidden in diffraction patterns computed from $P\bar{3}$ space group, i.e., they have a three orders of magnitude lower scattering factor compared to $P31m$, because of its lower symmetry.

As the sample is better described in $P31m$ space group structural investigation of both pristine and Rh-substituted samples was carried out employing this model.
It is worth noting that $P31m$ space group is not centrosymmetric, which makes it a theoretically proposed candidate for realization of Weyl states~\cite{yao2016} and an attractive system for studying the triply degenerate point fermions in that system~\cite{Gao2018}.

\begin{table}
  \caption{\label{refine-table}
    Structural parameters and residual factors of Rietveld analysis}
  \begin{tabular}{lrrr}
    \toprule{}
    \multirow{2}{*}{Parameter}     & \multicolumn{3}{c}{Composition, \ce{Pt_{1-x}Rh_{x}Bi2}, x}                            \\
    \cline{2-4}
                                   & \multicolumn{2}{c}{0}         & 0.03                                                  \\
    \midrule{}
    Wavelength (\AA)               & 1.78996                       & 1.78996                   & 1.78996                   \\
    $2\theta$  range (\textdegree) & 10--111.955                   & 10--111.955               & 10--111.995               \\
    Step Size (\textdegree)        & 0.015                         & 0.015                     & 0.015                     \\
    Temperature ($K$)              & 293                           & 293                       & 293                       \\
    Space Group                    & $P\bar{3}$ (\textnumero{}147) & $P31m$ (\textnumero{}157) & $P31m$ (\textnumero{}157) \\
    $a$ (\AA)                      & 6.5731(8)                     & 6.5731(6)                 & 6.57696(2)                \\
    $c$ (\AA)                      & 6.162(2)                      & 6.1619(13)                & 6.14796(4)                \\
    $U_{isotropic}$:               &                               &                           &                           \\
    $U_{Pt1}$                      & -0.0857(14)                   & 0.025(1)                  & 0.0077(5)                 \\
    $U_{Pt2}$                      & 0.043(3)                      & n/a                       & n/a                       \\
    $U_{Bi1}$                      & -0.0183(9)                    & 0.052(2)                  & 0.0164(9)                 \\
    $U_{Bi2}$                      & n/a                           & 0.0209(14)                & 0.0087(5)                 \\
    $U_{Bi3}$                      & n/a                           & 0.0218(10)                & 0.0142(5)                 \\
    $R$                            & 0.1163                        & 0.0520                    & 0.0226                    \\
    $wR$                           & 0.1892                        & 0.0725                    & 0.0324                    \\
    Goodness-of-Fit                & 8.45                          & 6.68                      & 1.92                      \\
    \bottomrule{}
  \end{tabular}
\end{table}

\begin{table}
  \caption{\label{wyckoff}
    Refined atomic coordinates for \ce{Pt_{1-x}Rh_xBi2}}
  \begin{tabularx}{\columnwidth}{Xllrrr}
    \toprule{}
    Sample                          & Atom & site & $x$       & $y$   & $z$       \\
    \midrule{}
    \multirow{4}{*}{\ce{PtBi2}}           & Pt (Pt1)          & $3c$         & 0.2619(5) & 0     & 0.363(13) \\
                                    & Bi (Bi1)          & $1a$         & 0         & 0     & 0         \\
                                    & Bi (Bi2)          & $2b$         & $2/3$     & $1/3$ & 0.155(13) \\
                                    & Bi (Bi3)          & $3c$         & 0.6144(5) & 0     & 0.630(13) \\
    \midrule{}
    \multirow{4}{*}{\ce{Pt_{0.97}Rh_{0.03}Bi2}} & Pt/Rh (Pt1)       & $3c$         & 0.2617(2) & 0     & 0.3578(6) \\
                                    & Bi (Bi1)          & $1a$         & 0         & 0     & 0.0139(6) \\
                                    & Bi (Bi2)          & $2b$         & $2/3$     & $1/3$ & 0.1413(6) \\
                                    & Bi (Bi3)          & $3c$         & 0.6093(2) & 0     & 0.6345(5) \\
    \bottomrule{}
  \end{tabularx}
\end{table}

Results of the refinement of 3\% substituted sample are presented on the fig.~\ref{riet} and in the tables~\ref{refine-table} and~\ref{wyckoff}.
In case of the 3\% substitution refinement of Rh/Pt occupational parameters from powder data is not feasible due to low Rh content, so in the refinement model the ``Pt1'' position was set to be fully occupied by platinum.
With the substitution of \ce{Pt} by \ce{Rh} we observe a slight increase in $a$ parameter by $\Delta{}a\approx{}0.003$\,\AA{} and a noticeable decrease of parameter $c$ by $\Delta{}c\approx{}-0.02$\,\AA{}.
This effect might be another indication of solid solution formation and can be explained by compression of distorted Bi-octahedra, and, as a result, slight expansion of the Pt-Bi framework in the $ab$ plane.
This lattice deformation might be a helpful tool to study Weyl point behavior, since the position of such nodes in electronic structure is quite sensitive to changes in lattice parameters. Such lattice deformation might provide effect analogous to strain which, according to theoretical models, can result in experimentally measurable effects~\cite{PhysRevB.94.241405}, and it already was shown experimentally that substitution can enhance superconductivity in the sample~\cite{doi:10.1063/1.4947433}.

\FloatBarrier{}

\subsection{Resistivity}

\begin{figure*}
  \includegraphics[width=0.8\textwidth{}]{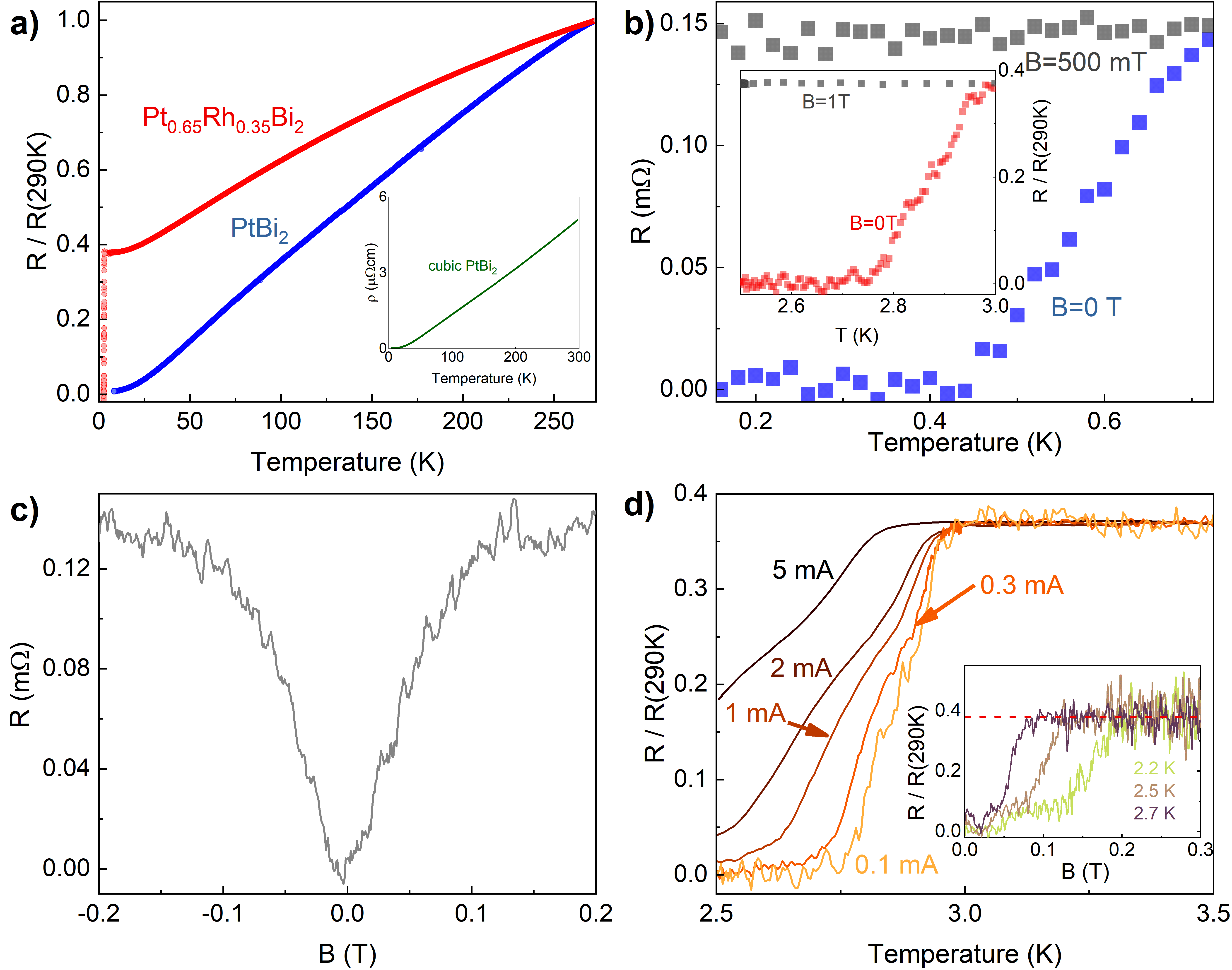}
  \caption{\label{img:res:trig}
    (a) T-dependence of the normalized resistivity $\rho{}/\rho{}_{290K}$ for
    trigonal \ce{PtBi2} and \ce{Pt_{0.65}Rh_{0.35}Bi2}, inset: T-dependence of
    the resistivity for cubic \ce{PtBi2} (b) T-dependence of $R$ for \ce{PtBi2}
    with magnetic field $B=0$\,T and $B=500$\,mT, applied parallel to the
    c-axis. Inset: T-dependence of $\rho{}/\rho{}_{290K}$ for
    \ce{Pt_{0.65}Rh_{0.35}Bi2} with magnetic field $B=0$\,T and $B=1$\,T, applied
    parallel to the c-axis.  (c) B-dependence of $R$ for \ce{PtBi2} at
    $T=100$~mK.  (d) Low temperature $\rho{}/\rho{}_{290K}$ vs T curves for
    \ce{Pt_{0.65}Rh_{0.35}Bi2} with different applied current from 0.1\,mA to
    5\,mA. Inset: magnetic field dependence of $\rho{}/\rho{}_{290K}$ for
    \ce{Pt_{0.65}Rh_{0.35}Bi2} at different $T = 2.2, 2.5$ and 2.7\,K with I =
    0.1\,mA.
  }
\end{figure*}

Inset of fig.~\ref{img:res:trig}a presents metallic nature of resistivity as a function of temperature in cubic \ce{PtBi2} with estimated RRR=650.
fig.~\ref{img:res:trig}a shows the temperature dependence of the normalized in-plane resistivity $\rho{}/\rho{}_{290K}$ of the \ce{Pt_{1-x}Rh_xBi2} crystals with x=0 and x=0.35.
\ce{PtBi2} presents a metallic behavior with a residual resistivity ratio (RRR) up to 132, evidencing the high purity of the sample.
\ce{Pt_{0.65}Rh_{0.35}Bi2} is also metallic but its RRR decreases to 2.7, due to the disorder introduced by the Rh substitution.

Measurements at very low temperature show a broad superconducting transition at 600\,mK for a current of $500\mu{}$A (fig.~\ref{img:res:trig}b).
This superconducting transition disappears in the presence of a 500\,mT magnetic field (fig.~\ref{img:res:trig}b) and we measure a critical magnetic field $B_c$ (defined as $R(Bc)=R_N/2$, $R_N$ being the resistance in the normal state) of 60\,mT (fig.~\ref{img:res:trig}c).
Further details regarding superconductivity anisotropy will be reported elsewhere~\cite{inprep}.
Similar transitions are observed in the \ce{Pt_{0.65}Rh_{0.35}Bi2} doped crystals but with a significantly larger critical temperature of 2.75\,K for a current of 0.1\,mA (see fig.~\ref{img:res:trig}a and inset in fig.~\ref{img:res:trig}b).
Again, a magnetic field of 1\,T aligned along the c-axis at a temperature of 1\,K suppresses this superconducting transition (inset of fig.~\ref{img:res:trig}b).
By increasing the current the transition systematically broadens (fig.~\ref{img:res:trig}d for \ce{Pt_{0.65}Rh_{0.35}Bi_2}), consistently with a progressive suppression of the superconducting phase (the not-well-defined geometry of the sample did not allow a reliable estimation of the critical current).
As expected for the superconducting state, the superconductivity is weakened by increasing the temperature and  the critical field decreases accordingly.
In the inset of fig.~\ref{img:res:trig}d the field-dependence of $\rho{}/\rho{}_{290K}$ is also presented: with increasing the temperature, the critical field, required to suppress the superconducting phase, diminishes as expected.

\subsection{Magnetization}

\begin{figure}
  \centering
  \includegraphics[width=\columnwidth{}]{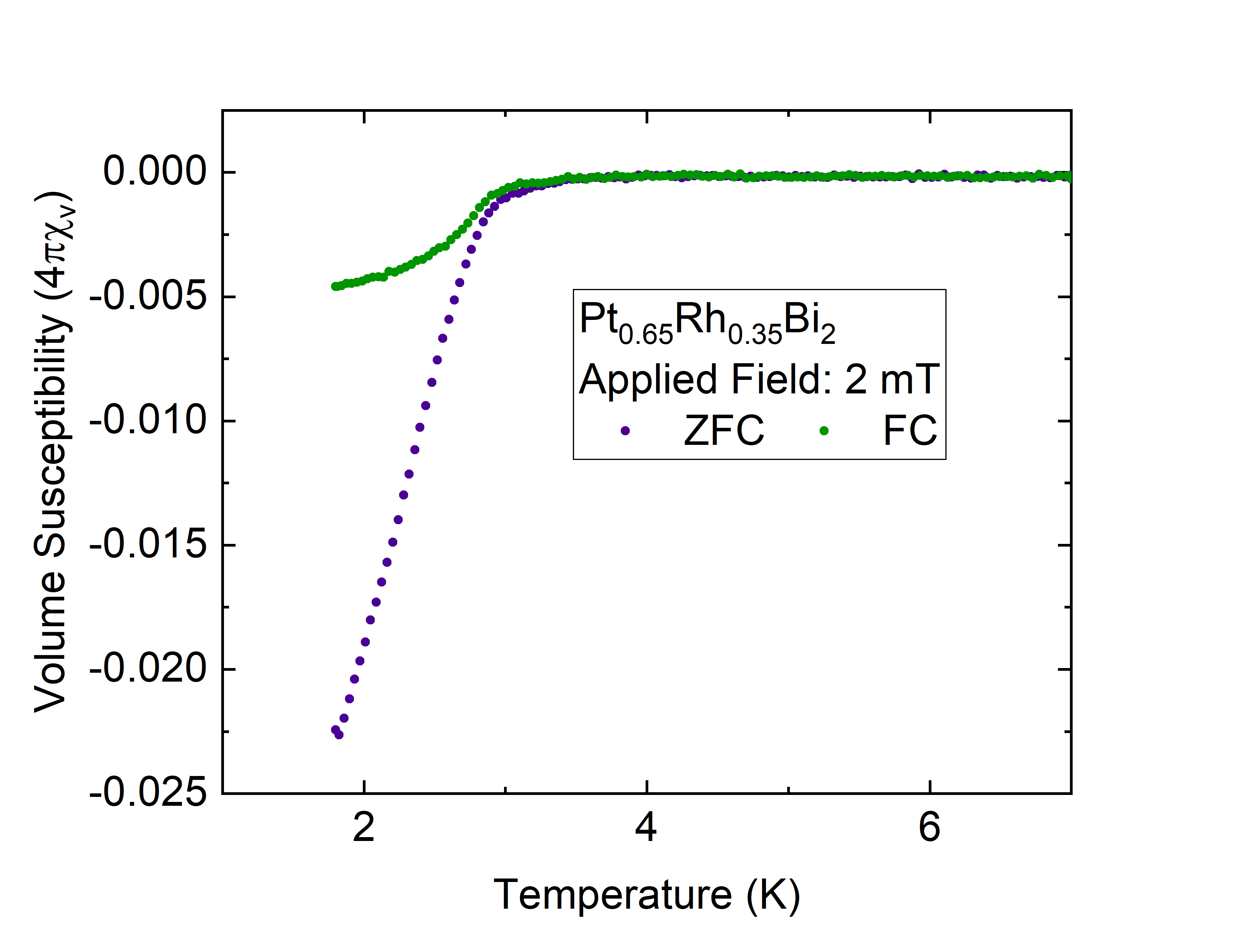}
  \caption{\label{img:SQUID}
    Temperature dependence of volume susceptibility for \ce{Pt_{0.65}Rh_{0.35}Bi2} in applied field of $H||ab=2$\,mT.}
\end{figure}

Magnetization measurements in the temperature range of $T=$1.8--300\,K in 0.5\,T field show diamagnetic behavior for both parent and Rh-substituted compounds with a Curie tail region at low temperatures, perhaps due to some paramagnetic impurities.
fig.~\ref{img:SQUID} presents the temperature dependent volume susceptibility~($\chi_{\text{vol}}$) for the \ce{Pt_{0.65}Rh_{0.35}Bi2} compound.
$\chi_{\text{vol}}$ was deduced from the measured magnetization vs temperature dependence and has not been corrected for demagnetization effects.
The sharp onset of the superconducting transition starts at $T_c\approx{}3$\,K.
However, the saturation is not seen down to 1.8\,K, probably due to temperature limitations of the device.
Observed $T_c$ is in line with the $T_c$ estimated from resistivity measurements.

\subsection{Electronic band structure calculation}

\begin{figure*}[h]
  \includegraphics[width=0.8\textwidth]{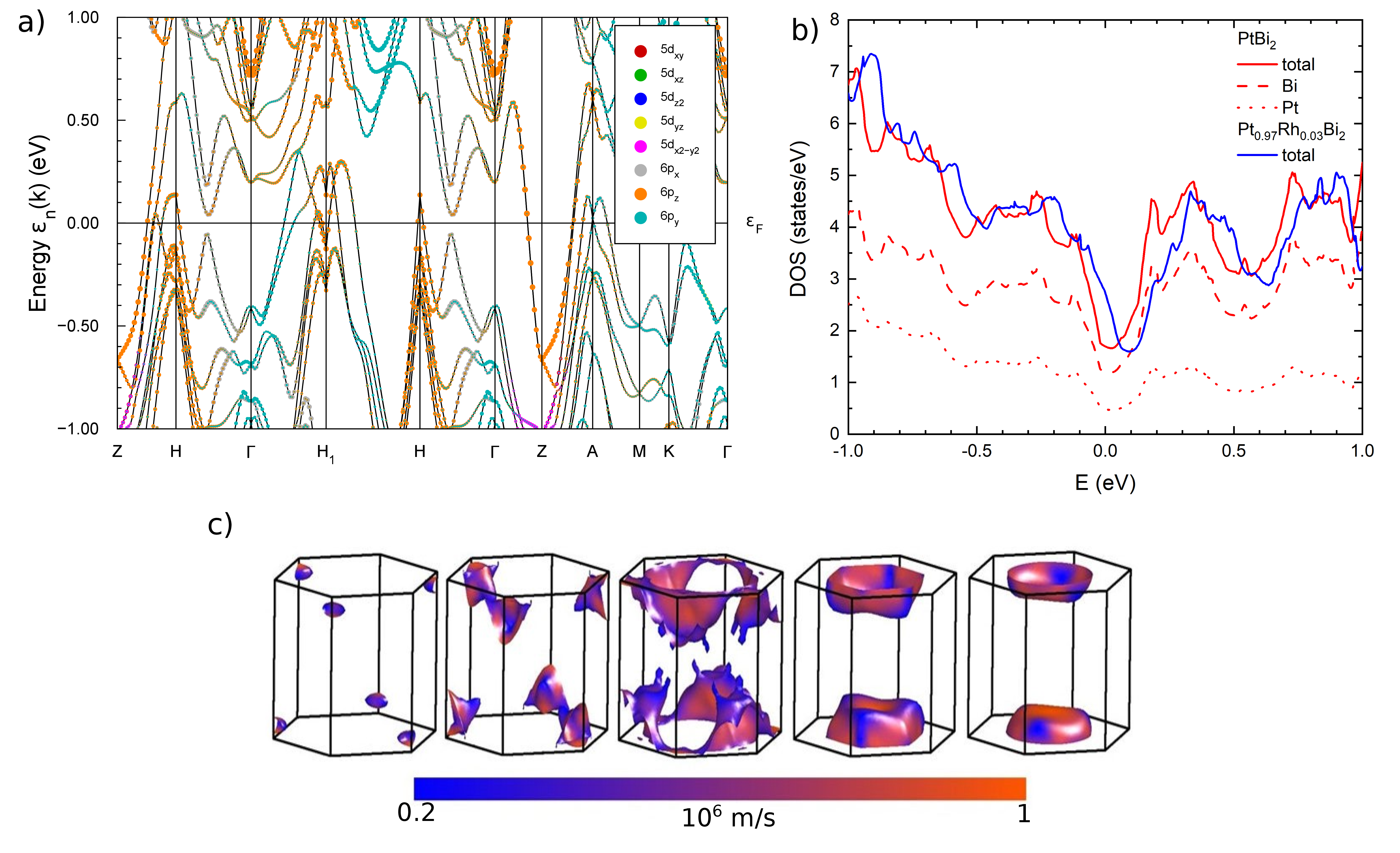}
  \caption{\label{theory}
    (a) Orbital decomposed band structure of trigonal \ce{PtBi2}, showing the contribution of 5d Pt and 6p Bi orbitals.
    (b) Total DOS for trigonal \ce{PtBi2} and partial contribution of Pt and Bi orbitals and DOS in \ce{Pt_{0.97}Rh_{0.03}Bi2}.
    (c) Fermi surface of trigonal \ce{PtBi2}}
\end{figure*}

Fig.~\ref{theory}b shows electronic density of states (DOS).
The density of states (DOS) shows a minimum for the stochiometric compound at the Fermi level.
Close to the Fermi level, only 6p Bi and 5d Pt states are present.
The orbital projected band structure is presented at fig.~\ref{theory}a and correspondent Fermi surface is shown in fig.~\ref{theory}c.
The colormap in the figure shows the velocity of the corresponding groups of electrons at the Fermi level.
The obtained band structure agrees with one reported earlier~\cite{Gao2018}.
The substitution of Pt by Rh leads to hole doping and enhancement of the DOS at the Fermi level.
The rise of superconducting critical temperature with Rh doping may be attributes this DOS enhancement.

\section{Conclusion}

In summary, we have successfully grown single crystals of both trigonal and pyrite-type polymorphic modifications of \ce{PtBi2}, as well as trigonal \ce{Pt_{1-x}Rh_xBi2} for x=0.03, 0.35 via self-flux technique.
As grown crystals were carefully characterized by SEM/EDX, powder X-ray diffraction and SAED.
The pyrite type crystallizes in \ce{FeS2} structure type with space group $Pa\bar{3}$, where trigonal modification crystallizes in $P31m$ space group as opposed to $P\bar{3}$.
Further, we have successfully grown the single crystals of Rh-doped \ce{PtBi2} in trigonal modification, which also shows \ce{Pt_{1-x}Rh_xBi2} solid solution formation.
Structural characterization demonstrates that crystal structure is preserved up to at least $x=0.35$.
Resistivity measurements for as grown crystals show metallic nature.
For pristine trigonal \ce{PtBi2} superconducting transition is found at 600\,mK.
For $x=0.35$ compound as measured superconducting transition temperature is enhanced up to 2.7\,K compared to pristine \ce{PtBi2} from both resistivity as well as from susceptibility measurements, which is in line with DOS shift near the Fermi level according to calculations.
Our findings together with data published previously make \ce{PtBi2}-family of materials a strong candidate for topological superconductivity.
This findings warrant further research by techniques such as quantum transport, scanning tunneling microscopy, as well as
the effect of the substitution on the non-trivial band structure of the compound has to be further investigated by revisiting electronic structure measurements by ARPES\@.

\subsection*{Acknowledgments}

SA acknowledges support of Deutsche Forschungsgemeinschaft (DFG) through Grant \textnumero{}AS 523/4--1.
SA, DVE \& BB also acknowledge support of DFG through Projekt \textnumero{}405940956.
IK and BB acknowledge support of BMBF through UKRATOP (BMBF), FKZ: 01DK18002.
JD acknowledges financial support by the Deutsche Forschungsgemeinschaft (DFG) through SPP 1666 Topological Insulators program and the Würzburg-Dresden Cluster of Excellence on Complexity and Topology in Quantum Matter --- ct.qmat (EXC 2147, project-id 39085490).
AL, DW and SS acknowledge funding from DFG SFB 1415 --- project-ID 417590517.
This project has received funding from the European Research Council (ERC) under the European Union’s Horizon 2020 research and innovation programme (Grant Agreement \textnumero{}647276-MARS-ERC-2014-CoG).

\subsection*{Data Availability}
The datasets analyzed during the current study are available from the corresponding authors upon reasonable request.

\subsection*{Competing Interests}
Authors declare no competing financial or non-financial interests.

\bibliography{PtBi2}

\begin{thebibliography}{33}%
\makeatletter
\providecommand \@ifxundefined [1]{%
 \@ifx{#1\undefined}
}%
\providecommand \@ifnum [1]{%
 \ifnum #1\expandafter \@firstoftwo
 \else \expandafter \@secondoftwo
 \fi
}%
\providecommand \@ifx [1]{%
 \ifx #1\expandafter \@firstoftwo
 \else \expandafter \@secondoftwo
 \fi
}%
\providecommand \natexlab [1]{#1}%
\providecommand \enquote  [1]{``#1''}%
\providecommand \bibnamefont  [1]{#1}%
\providecommand \bibfnamefont [1]{#1}%
\providecommand \citenamefont [1]{#1}%
\providecommand \href@noop [0]{\@secondoftwo}%
\providecommand \href [0]{\begingroup \@sanitize@url \@href}%
\providecommand \@href[1]{\@@startlink{#1}\@@href}%
\providecommand \@@href[1]{\endgroup#1\@@endlink}%
\providecommand \@sanitize@url [0]{\catcode `\\12\catcode `\$12\catcode
  `\&12\catcode `\#12\catcode `\^12\catcode `\_12\catcode `\%12\relax}%
\providecommand \@@startlink[1]{}%
\providecommand \@@endlink[0]{}%
\providecommand \url  [0]{\begingroup\@sanitize@url \@url }%
\providecommand \@url [1]{\endgroup\@href {#1}{\urlprefix }}%
\providecommand \urlprefix  [0]{URL }%
\providecommand \Eprint [0]{\href }%
\providecommand \doibase [0]{https://doi.org/}%
\providecommand \selectlanguage [0]{\@gobble}%
\providecommand \bibinfo  [0]{\@secondoftwo}%
\providecommand \bibfield  [0]{\@secondoftwo}%
\providecommand \translation [1]{[#1]}%
\providecommand \BibitemOpen [0]{}%
\providecommand \bibitemStop [0]{}%
\providecommand \bibitemNoStop [0]{.\EOS\space}%
\providecommand \EOS [0]{\spacefactor3000\relax}%
\providecommand \BibitemShut  [1]{\csname bibitem#1\endcsname}%
\let\auto@bib@innerbib\@empty
\bibitem [{\citenamefont {Lee}\ \emph {et~al.}(2019)\citenamefont {Lee},
  \citenamefont {Sharma}, \citenamefont {Lima-Sharma}, \citenamefont {Pan},\
  and\ \citenamefont {Nenoff}}]{lee2019topological}%
  \BibitemOpen
  \bibfield  {author} {\bibinfo {author} {\bibfnamefont {S.~R.}\ \bibnamefont
  {Lee}}, \bibinfo {author} {\bibfnamefont {P.~A.}\ \bibnamefont {Sharma}},
  \bibinfo {author} {\bibfnamefont {A.~L.}\ \bibnamefont {Lima-Sharma}},
  \bibinfo {author} {\bibfnamefont {W.}~\bibnamefont {Pan}},\ and\ \bibinfo
  {author} {\bibfnamefont {T.~M.}\ \bibnamefont {Nenoff}},\ }\bibfield  {title}
  {\bibinfo {title} {Topological quantum materials for realizing majorana
  quasiparticles},\ }\href@noop {} {\bibfield  {journal} {\bibinfo  {journal}
  {Chemistry of Materials}\ }\textbf {\bibinfo {volume} {31}},\ \bibinfo
  {pages} {26} (\bibinfo {year} {2019})}\BibitemShut {NoStop}%
\bibitem [{\citenamefont {Savage}(2018)}]{savage2018topology}%
  \BibitemOpen
  \bibfield  {author} {\bibinfo {author} {\bibfnamefont {N.}~\bibnamefont
  {Savage}},\ }\bibfield  {title} {\bibinfo {title} {Topology shapes a search
  for new materials},\ }\href@noop {} {\bibfield  {journal} {\bibinfo
  {journal} {ACS Central Science}\ }\textbf {\bibinfo {volume} {4}},\ \bibinfo
  {pages} {523} (\bibinfo {year} {2018})}\BibitemShut {NoStop}%
\bibitem [{\citenamefont {Yan}\ and\ \citenamefont
  {Felser}(2017)}]{yan2017topological}%
  \BibitemOpen
  \bibfield  {author} {\bibinfo {author} {\bibfnamefont {B.}~\bibnamefont
  {Yan}}\ and\ \bibinfo {author} {\bibfnamefont {C.}~\bibnamefont {Felser}},\
  }\bibfield  {title} {\bibinfo {title} {Topological materials: Weyl
  semimetals},\ }\href@noop {} {\bibfield  {journal} {\bibinfo  {journal}
  {Annual Review of Condensed Matter Physics}\ }\textbf {\bibinfo {volume}
  {8}},\ \bibinfo {pages} {337} (\bibinfo {year} {2017})}\BibitemShut {NoStop}%
\bibitem [{\citenamefont {Xu}\ \emph {et~al.}(2015)\citenamefont {Xu},
  \citenamefont {Belopolski}, \citenamefont {Alidoust}, \citenamefont
  {Neupane}, \citenamefont {Bian}, \citenamefont {Zhang}, \citenamefont
  {Sankar}, \citenamefont {Chang}, \citenamefont {Yuan}, \citenamefont {Lee}
  \emph {et~al.}}]{xu2015discovery}%
  \BibitemOpen
  \bibfield  {author} {\bibinfo {author} {\bibfnamefont {S.-Y.}\ \bibnamefont
  {Xu}}, \bibinfo {author} {\bibfnamefont {I.}~\bibnamefont {Belopolski}},
  \bibinfo {author} {\bibfnamefont {N.}~\bibnamefont {Alidoust}}, \bibinfo
  {author} {\bibfnamefont {M.}~\bibnamefont {Neupane}}, \bibinfo {author}
  {\bibfnamefont {G.}~\bibnamefont {Bian}}, \bibinfo {author} {\bibfnamefont
  {C.}~\bibnamefont {Zhang}}, \bibinfo {author} {\bibfnamefont
  {R.}~\bibnamefont {Sankar}}, \bibinfo {author} {\bibfnamefont
  {G.}~\bibnamefont {Chang}}, \bibinfo {author} {\bibfnamefont
  {Z.}~\bibnamefont {Yuan}}, \bibinfo {author} {\bibfnamefont {C.-C.}\
  \bibnamefont {Lee}}, \emph {et~al.},\ }\bibfield  {title} {\bibinfo {title}
  {Discovery of a weyl fermion semimetal and topological fermi arcs},\
  }\href@noop {} {\bibfield  {journal} {\bibinfo  {journal} {Science}\ }\textbf
  {\bibinfo {volume} {349}},\ \bibinfo {pages} {613} (\bibinfo {year}
  {2015})}\BibitemShut {NoStop}%
\bibitem [{\citenamefont {Borisenko}\ \emph {et~al.}(2014)\citenamefont
  {Borisenko}, \citenamefont {Gibson}, \citenamefont {Evtushinsky},
  \citenamefont {Zabolotnyy}, \citenamefont {B{\"u}chner},\ and\ \citenamefont
  {Cava}}]{borisenko2014experimental}%
  \BibitemOpen
  \bibfield  {author} {\bibinfo {author} {\bibfnamefont {S.}~\bibnamefont
  {Borisenko}}, \bibinfo {author} {\bibfnamefont {Q.}~\bibnamefont {Gibson}},
  \bibinfo {author} {\bibfnamefont {D.}~\bibnamefont {Evtushinsky}}, \bibinfo
  {author} {\bibfnamefont {V.}~\bibnamefont {Zabolotnyy}}, \bibinfo {author}
  {\bibfnamefont {B.}~\bibnamefont {B{\"u}chner}},\ and\ \bibinfo {author}
  {\bibfnamefont {R.~J.}\ \bibnamefont {Cava}},\ }\bibfield  {title} {\bibinfo
  {title} {Experimental realization of a three-dimensional dirac semimetal},\
  }\href@noop {} {\bibfield  {journal} {\bibinfo  {journal} {Physical review
  letters}\ }\textbf {\bibinfo {volume} {113}},\ \bibinfo {pages} {027603}
  (\bibinfo {year} {2014})}\BibitemShut {NoStop}%
\bibitem [{\citenamefont {Shekhar}\ \emph {et~al.}(2015)\citenamefont
  {Shekhar}, \citenamefont {Nayak}, \citenamefont {Sun}, \citenamefont
  {Schmidt}, \citenamefont {Nicklas}, \citenamefont {Leermakers}, \citenamefont
  {Zeitler}, \citenamefont {Skourski}, \citenamefont {Wosnitza}, \citenamefont
  {Liu} \emph {et~al.}}]{shekhar2015extremely}%
  \BibitemOpen
  \bibfield  {author} {\bibinfo {author} {\bibfnamefont {C.}~\bibnamefont
  {Shekhar}}, \bibinfo {author} {\bibfnamefont {A.~K.}\ \bibnamefont {Nayak}},
  \bibinfo {author} {\bibfnamefont {Y.}~\bibnamefont {Sun}}, \bibinfo {author}
  {\bibfnamefont {M.}~\bibnamefont {Schmidt}}, \bibinfo {author} {\bibfnamefont
  {M.}~\bibnamefont {Nicklas}}, \bibinfo {author} {\bibfnamefont
  {I.}~\bibnamefont {Leermakers}}, \bibinfo {author} {\bibfnamefont
  {U.}~\bibnamefont {Zeitler}}, \bibinfo {author} {\bibfnamefont
  {Y.}~\bibnamefont {Skourski}}, \bibinfo {author} {\bibfnamefont
  {J.}~\bibnamefont {Wosnitza}}, \bibinfo {author} {\bibfnamefont
  {Z.}~\bibnamefont {Liu}}, \emph {et~al.},\ }\bibfield  {title} {\bibinfo
  {title} {Extremely large magnetoresistance and ultrahigh mobility in the
  topological weyl semimetal candidate {NbP}},\ }\href@noop {} {\bibfield
  {journal} {\bibinfo  {journal} {Nature Physics}\ }\textbf {\bibinfo {volume}
  {11}},\ \bibinfo {pages} {645} (\bibinfo {year} {2015})}\BibitemShut
  {NoStop}%
\bibitem [{\citenamefont {Liang}\ \emph {et~al.}(2015)\citenamefont {Liang},
  \citenamefont {Gibson}, \citenamefont {Ali}, \citenamefont {Liu},
  \citenamefont {Cava},\ and\ \citenamefont {Ong}}]{liang2015ultrahigh}%
  \BibitemOpen
  \bibfield  {author} {\bibinfo {author} {\bibfnamefont {T.}~\bibnamefont
  {Liang}}, \bibinfo {author} {\bibfnamefont {Q.}~\bibnamefont {Gibson}},
  \bibinfo {author} {\bibfnamefont {M.~N.}\ \bibnamefont {Ali}}, \bibinfo
  {author} {\bibfnamefont {M.}~\bibnamefont {Liu}}, \bibinfo {author}
  {\bibfnamefont {R.~J.}\ \bibnamefont {Cava}},\ and\ \bibinfo {author}
  {\bibfnamefont {N.~P.}\ \bibnamefont {Ong}},\ }\bibfield  {title} {\bibinfo
  {title} {Ultrahigh mobility and giant magnetoresistance in the dirac
  semimetal {Cd$_3$As$_2$}},\ }\href@noop {} {\bibfield  {journal} {\bibinfo
  {journal} {Nature materials}\ }\textbf {\bibinfo {volume} {14}},\ \bibinfo
  {pages} {280} (\bibinfo {year} {2015})}\BibitemShut {NoStop}%
\bibitem [{\citenamefont {Arnold}\ \emph {et~al.}(2016)\citenamefont {Arnold},
  \citenamefont {Shekhar}, \citenamefont {Wu}, \citenamefont {Sun},
  \citenamefont {Dos~Reis}, \citenamefont {Kumar}, \citenamefont {Naumann},
  \citenamefont {Ajeesh}, \citenamefont {Schmidt}, \citenamefont {Grushin}
  \emph {et~al.}}]{arnold2016negative}%
  \BibitemOpen
  \bibfield  {author} {\bibinfo {author} {\bibfnamefont {F.}~\bibnamefont
  {Arnold}}, \bibinfo {author} {\bibfnamefont {C.}~\bibnamefont {Shekhar}},
  \bibinfo {author} {\bibfnamefont {S.-C.}\ \bibnamefont {Wu}}, \bibinfo
  {author} {\bibfnamefont {Y.}~\bibnamefont {Sun}}, \bibinfo {author}
  {\bibfnamefont {R.~D.}\ \bibnamefont {Dos~Reis}}, \bibinfo {author}
  {\bibfnamefont {N.}~\bibnamefont {Kumar}}, \bibinfo {author} {\bibfnamefont
  {M.}~\bibnamefont {Naumann}}, \bibinfo {author} {\bibfnamefont {M.~O.}\
  \bibnamefont {Ajeesh}}, \bibinfo {author} {\bibfnamefont {M.}~\bibnamefont
  {Schmidt}}, \bibinfo {author} {\bibfnamefont {A.~G.}\ \bibnamefont
  {Grushin}}, \emph {et~al.},\ }\bibfield  {title} {\bibinfo {title} {Negative
  magnetoresistance without well-defined chirality in the weyl semimetal
  {TaP}},\ }\href@noop {} {\bibfield  {journal} {\bibinfo  {journal} {Nature
  communications}\ }\textbf {\bibinfo {volume} {7}},\ \bibinfo {pages} {11615}
  (\bibinfo {year} {2016})}\BibitemShut {NoStop}%
\bibitem [{\citenamefont {Zyuzin}\ and\ \citenamefont
  {Tiwari}(2016)}]{zyuzin2016intrinsic}%
  \BibitemOpen
  \bibfield  {author} {\bibinfo {author} {\bibfnamefont {A.~A.}\ \bibnamefont
  {Zyuzin}}\ and\ \bibinfo {author} {\bibfnamefont {R.~P.}\ \bibnamefont
  {Tiwari}},\ }\bibfield  {title} {\bibinfo {title} {Intrinsic anomalous hall
  effect in {type-II} {Weyl} semimetals},\ }\href@noop {} {\bibfield  {journal}
  {\bibinfo  {journal} {JETP letters}\ }\textbf {\bibinfo {volume} {103}},\
  \bibinfo {pages} {717} (\bibinfo {year} {2016})}\BibitemShut {NoStop}%
\bibitem [{\citenamefont {Vaezi}(2013)}]{vaezi2013fractional}%
  \BibitemOpen
  \bibfield  {author} {\bibinfo {author} {\bibfnamefont {A.}~\bibnamefont
  {Vaezi}},\ }\bibfield  {title} {\bibinfo {title} {Fractional topological
  superconductor with fractionalized majorana fermions},\ }\href@noop {}
  {\bibfield  {journal} {\bibinfo  {journal} {Physical Review B}\ }\textbf
  {\bibinfo {volume} {87}},\ \bibinfo {pages} {035132} (\bibinfo {year}
  {2013})}\BibitemShut {NoStop}%
\bibitem [{\citenamefont {{Alekseevski, N. E.}}\ and\ \citenamefont {{Gaidukov,
  Yu. P.}}(1953)}]{alekseevski1953}%
  \BibitemOpen
  \bibfield  {author} {\bibinfo {author} {\bibnamefont {{Alekseevski, N. E.}}}\
  and\ \bibinfo {author} {\bibnamefont {{Gaidukov, Yu. P.}}},\ }\href@noop {}
  {\bibfield  {journal} {\bibinfo  {journal} {Zhurnal eksperimental'noi i
  teoreticheskoi fiziki}\ }\textbf {\bibinfo {volume} {25}},\ \bibinfo {pages}
  {383} (\bibinfo {year} {1953})}\BibitemShut {NoStop}%
\bibitem [{\citenamefont {Okamoto}(1991)}]{okamoto1991bi}%
  \BibitemOpen
  \bibfield  {author} {\bibinfo {author} {\bibfnamefont {H.}~\bibnamefont
  {Okamoto}},\ }\bibfield  {title} {\bibinfo {title} {The {Bi-Pt}
  (bismuth-platinum) system},\ }\href@noop {} {\bibfield  {journal} {\bibinfo
  {journal} {Journal of phase equilibria}\ }\textbf {\bibinfo {volume} {12}},\
  \bibinfo {pages} {207} (\bibinfo {year} {1991})}\BibitemShut {NoStop}%
\bibitem [{\citenamefont {Zhuravlev}\ and\ \citenamefont
  {Stepanova}(1962)}]{zhu1}%
  \BibitemOpen
  \bibfield  {author} {\bibinfo {author} {\bibfnamefont {N.}~\bibnamefont
  {Zhuravlev}}\ and\ \bibinfo {author} {\bibfnamefont {A.}~\bibnamefont
  {Stepanova}},\ }\bibfield  {title} {\bibinfo {title} {An x-ray investigation
  of superconducting alloys of bismuth with platinum in the range 20-640°c},\
  }\href@noop {} {\bibfield  {journal} {\bibinfo  {journal} {Soviet Physics -
  Crystallography}\ }\textbf {\bibinfo {volume} {7}},\ \bibinfo {pages} {241}
  (\bibinfo {year} {1962})}\BibitemShut {NoStop}%
\bibitem [{\citenamefont {Biswas}\ and\ \citenamefont {Schubert}(1969)}]{bar}%
  \BibitemOpen
  \bibfield  {author} {\bibinfo {author} {\bibfnamefont {T.}~\bibnamefont
  {Biswas}}\ and\ \bibinfo {author} {\bibfnamefont {K.}~\bibnamefont
  {Schubert}},\ }\bibfield  {title} {\bibinfo {title} {Strukturuntersuchungen
  in den mischungen {Pt-Tl-Pb} und {Pt-Pb-Bi}},\ }\href@noop {} {\bibfield
  {journal} {\bibinfo  {journal} {Journal of the Less Common Metals}\ }\textbf
  {\bibinfo {volume} {19}},\ \bibinfo {pages} {223 } (\bibinfo {year}
  {1969})}\BibitemShut {NoStop}%
\bibitem [{\citenamefont {Gao}\ \emph {et~al.}(2017)\citenamefont {Gao},
  \citenamefont {Hao}, \citenamefont {Zheng}, \citenamefont {Ning},
  \citenamefont {Wu}, \citenamefont {Zhu}, \citenamefont {Zheng}, \citenamefont
  {Zhang}, \citenamefont {Lu}, \citenamefont {Zhang} \emph
  {et~al.}}]{gao2017extremely}%
  \BibitemOpen
  \bibfield  {author} {\bibinfo {author} {\bibfnamefont {W.}~\bibnamefont
  {Gao}}, \bibinfo {author} {\bibfnamefont {N.}~\bibnamefont {Hao}}, \bibinfo
  {author} {\bibfnamefont {F.-W.}\ \bibnamefont {Zheng}}, \bibinfo {author}
  {\bibfnamefont {W.}~\bibnamefont {Ning}}, \bibinfo {author} {\bibfnamefont
  {M.}~\bibnamefont {Wu}}, \bibinfo {author} {\bibfnamefont {X.}~\bibnamefont
  {Zhu}}, \bibinfo {author} {\bibfnamefont {G.}~\bibnamefont {Zheng}}, \bibinfo
  {author} {\bibfnamefont {J.}~\bibnamefont {Zhang}}, \bibinfo {author}
  {\bibfnamefont {J.}~\bibnamefont {Lu}}, \bibinfo {author} {\bibfnamefont
  {H.}~\bibnamefont {Zhang}}, \emph {et~al.},\ }\bibfield  {title} {\bibinfo
  {title} {Extremely large magnetoresistance in a topological semimetal
  candidate pyrite {PtBi$_2$}},\ }\href@noop {} {\bibfield  {journal} {\bibinfo
   {journal} {Physical review letters}\ }\textbf {\bibinfo {volume} {118}},\
  \bibinfo {pages} {256601} (\bibinfo {year} {2017})}\BibitemShut {NoStop}%
\bibitem [{\citenamefont {Gibson}\ \emph {et~al.}(2015)\citenamefont {Gibson},
  \citenamefont {Schoop}, \citenamefont {Muechler}, \citenamefont {Xie},
  \citenamefont {Hirschberger}, \citenamefont {Ong}, \citenamefont {Car},\ and\
  \citenamefont {Cava}}]{PhysRevB.91.205128}%
  \BibitemOpen
  \bibfield  {author} {\bibinfo {author} {\bibfnamefont {Q.~D.}\ \bibnamefont
  {Gibson}}, \bibinfo {author} {\bibfnamefont {L.~M.}\ \bibnamefont {Schoop}},
  \bibinfo {author} {\bibfnamefont {L.}~\bibnamefont {Muechler}}, \bibinfo
  {author} {\bibfnamefont {L.~S.}\ \bibnamefont {Xie}}, \bibinfo {author}
  {\bibfnamefont {M.}~\bibnamefont {Hirschberger}}, \bibinfo {author}
  {\bibfnamefont {N.~P.}\ \bibnamefont {Ong}}, \bibinfo {author} {\bibfnamefont
  {R.}~\bibnamefont {Car}},\ and\ \bibinfo {author} {\bibfnamefont {R.~J.}\
  \bibnamefont {Cava}},\ }\bibfield  {title} {\bibinfo {title}
  {Three-dimensional dirac semimetals: Design principles and predictions of new
  materials},\ }\href@noop {} {\bibfield  {journal} {\bibinfo  {journal} {Phys.
  Rev. B}\ }\textbf {\bibinfo {volume} {91}},\ \bibinfo {pages} {205128}
  (\bibinfo {year} {2015})}\BibitemShut {NoStop}%
\bibitem [{\citenamefont {Chen}\ \emph {et~al.}(2018)\citenamefont {Chen},
  \citenamefont {Shao}, \citenamefont {Gu}, \citenamefont {Zhou}, \citenamefont
  {An}, \citenamefont {Zhou}, \citenamefont {Zhu}, \citenamefont {Chen},
  \citenamefont {Tian}, \citenamefont {Sun},\ and\ \citenamefont
  {Yang}}]{PhysRevMaterials.2.054203}%
  \BibitemOpen
  \bibfield  {author} {\bibinfo {author} {\bibfnamefont {X.}~\bibnamefont
  {Chen}}, \bibinfo {author} {\bibfnamefont {D.}~\bibnamefont {Shao}}, \bibinfo
  {author} {\bibfnamefont {C.}~\bibnamefont {Gu}}, \bibinfo {author}
  {\bibfnamefont {Y.}~\bibnamefont {Zhou}}, \bibinfo {author} {\bibfnamefont
  {C.}~\bibnamefont {An}}, \bibinfo {author} {\bibfnamefont {Y.}~\bibnamefont
  {Zhou}}, \bibinfo {author} {\bibfnamefont {X.}~\bibnamefont {Zhu}}, \bibinfo
  {author} {\bibfnamefont {T.}~\bibnamefont {Chen}}, \bibinfo {author}
  {\bibfnamefont {M.}~\bibnamefont {Tian}}, \bibinfo {author} {\bibfnamefont
  {J.}~\bibnamefont {Sun}},\ and\ \bibinfo {author} {\bibfnamefont
  {Z.}~\bibnamefont {Yang}},\ }\bibfield  {title} {\bibinfo {title}
  {Pressure-induced multiband superconductivity in pyrite
  {$\mathrm{PtB}{\mathrm{i}}_{2}$} with perfect electron-hole compensation},\
  }\href@noop {} {\bibfield  {journal} {\bibinfo  {journal} {Phys. Rev.
  Materials}\ }\textbf {\bibinfo {volume} {2}},\ \bibinfo {pages} {054203}
  (\bibinfo {year} {2018})}\BibitemShut {NoStop}%
\bibitem [{\citenamefont {{Thirupathaiah}}\ \emph {et~al.}(2020)\citenamefont
  {{Thirupathaiah}}, \citenamefont {{Kushnirenk}}, \citenamefont {{Koepernik}},
  \citenamefont {{Piening}}, \citenamefont {{Buechner}}, \citenamefont
  {{Aswartham}}, \citenamefont {{van den Brink}}, \citenamefont {{Borisenko}},\
  and\ \citenamefont {{Fulga}}}]{2020arXiv200608642T}%
  \BibitemOpen
  \bibfield  {author} {\bibinfo {author} {\bibfnamefont {S.}~\bibnamefont
  {{Thirupathaiah}}}, \bibinfo {author} {\bibfnamefont {Y.~S.}\ \bibnamefont
  {{Kushnirenk}}}, \bibinfo {author} {\bibfnamefont {K.}~\bibnamefont
  {{Koepernik}}}, \bibinfo {author} {\bibfnamefont {B.~R.}\ \bibnamefont
  {{Piening}}}, \bibinfo {author} {\bibfnamefont {B.}~\bibnamefont
  {{Buechner}}}, \bibinfo {author} {\bibfnamefont {S.}~\bibnamefont
  {{Aswartham}}}, \bibinfo {author} {\bibfnamefont {J.}~\bibnamefont {{van den
  Brink}}}, \bibinfo {author} {\bibfnamefont {S.~V.}\ \bibnamefont
  {{Borisenko}}},\ and\ \bibinfo {author} {\bibfnamefont {I.~C.}\ \bibnamefont
  {{Fulga}}},\ }\bibfield  {title} {\bibinfo {title} {{Sixfold fermion near the
  Fermi level in cubic {PtBi$_2$}}},\ }\href@noop {} {\bibfield  {journal}
  {\bibinfo  {journal} {arXiv e-prints}\ } (\bibinfo {year} {2020})},\ \Eprint
  {https://arxiv.org/abs/2006.08642} {arXiv:2006.08642 [cond-mat.mes-hall]}
  \BibitemShut {NoStop}%
\bibitem [{\citenamefont {Yang}\ \emph {et~al.}(2016)\citenamefont {Yang},
  \citenamefont {Bai}, \citenamefont {Wang}, \citenamefont {Li}, \citenamefont
  {Chen}, \citenamefont {Chen}, \citenamefont {Li}, \citenamefont {Feng},
  \citenamefont {Zheng},\ and\ \citenamefont {Xu}}]{triclinMR}%
  \BibitemOpen
  \bibfield  {author} {\bibinfo {author} {\bibfnamefont {X.}~\bibnamefont
  {Yang}}, \bibinfo {author} {\bibfnamefont {H.}~\bibnamefont {Bai}}, \bibinfo
  {author} {\bibfnamefont {Z.}~\bibnamefont {Wang}}, \bibinfo {author}
  {\bibfnamefont {Y.}~\bibnamefont {Li}}, \bibinfo {author} {\bibfnamefont
  {Q.}~\bibnamefont {Chen}}, \bibinfo {author} {\bibfnamefont {J.}~\bibnamefont
  {Chen}}, \bibinfo {author} {\bibfnamefont {Y.}~\bibnamefont {Li}}, \bibinfo
  {author} {\bibfnamefont {C.}~\bibnamefont {Feng}}, \bibinfo {author}
  {\bibfnamefont {Y.}~\bibnamefont {Zheng}},\ and\ \bibinfo {author}
  {\bibfnamefont {Z.-a.}\ \bibnamefont {Xu}},\ }\bibfield  {title} {\bibinfo
  {title} {Giant linear magneto-resistance in nonmagnetic {PtBi$_2$}},\
  }\href@noop {} {\bibfield  {journal} {\bibinfo  {journal} {Applied Physics
  Letters}\ }\textbf {\bibinfo {volume} {108}},\ \bibinfo {pages} {252401}
  (\bibinfo {year} {2016})}\BibitemShut {NoStop}%
\bibitem [{\citenamefont {Thirupathaiah}\ \emph {et~al.}(2018)\citenamefont
  {Thirupathaiah}, \citenamefont {Kushnirenko}, \citenamefont {Haubold},
  \citenamefont {Fedorov}, \citenamefont {Rienks}, \citenamefont {Kim},
  \citenamefont {Yaresko}, \citenamefont {Blum}, \citenamefont {Aswartham},
  \citenamefont {B\"uchner},\ and\ \citenamefont
  {Borisenko}}]{thirupathaiah2018}%
  \BibitemOpen
  \bibfield  {author} {\bibinfo {author} {\bibfnamefont {S.}~\bibnamefont
  {Thirupathaiah}}, \bibinfo {author} {\bibfnamefont {Y.}~\bibnamefont
  {Kushnirenko}}, \bibinfo {author} {\bibfnamefont {E.}~\bibnamefont
  {Haubold}}, \bibinfo {author} {\bibfnamefont {A.~V.}\ \bibnamefont
  {Fedorov}}, \bibinfo {author} {\bibfnamefont {E.~D.~L.}\ \bibnamefont
  {Rienks}}, \bibinfo {author} {\bibfnamefont {T.~K.}\ \bibnamefont {Kim}},
  \bibinfo {author} {\bibfnamefont {A.~N.}\ \bibnamefont {Yaresko}}, \bibinfo
  {author} {\bibfnamefont {C.~G.~F.}\ \bibnamefont {Blum}}, \bibinfo {author}
  {\bibfnamefont {S.}~\bibnamefont {Aswartham}}, \bibinfo {author}
  {\bibfnamefont {B.}~\bibnamefont {B\"uchner}},\ and\ \bibinfo {author}
  {\bibfnamefont {S.~V.}\ \bibnamefont {Borisenko}},\ }\bibfield  {title}
  {\bibinfo {title} {Possible origin of linear magnetoresistance: Observation
  of dirac surface states in layered {${\mathrm{PtBi}}_{2}$}},\ }\href@noop {}
  {\bibfield  {journal} {\bibinfo  {journal} {Phys. Rev. B}\ }\textbf {\bibinfo
  {volume} {97}},\ \bibinfo {pages} {035133} (\bibinfo {year}
  {2018})}\BibitemShut {NoStop}%
\bibitem [{\citenamefont {Yao}\ \emph {et~al.}(2016)\citenamefont {Yao},
  \citenamefont {Du}, \citenamefont {Yang}, \citenamefont {Zheng},
  \citenamefont {Xu}, \citenamefont {Niu}, \citenamefont {Shen}, \citenamefont
  {Yang}, \citenamefont {Dudin}, \citenamefont {Kim}, \citenamefont {Hoesch},
  \citenamefont {Vobornik}, \citenamefont {Xu}, \citenamefont {Wan},
  \citenamefont {Feng},\ and\ \citenamefont {Shen}}]{yao2016}%
  \BibitemOpen
  \bibfield  {author} {\bibinfo {author} {\bibfnamefont {Q.}~\bibnamefont
  {Yao}}, \bibinfo {author} {\bibfnamefont {Y.~P.}\ \bibnamefont {Du}},
  \bibinfo {author} {\bibfnamefont {X.~J.}\ \bibnamefont {Yang}}, \bibinfo
  {author} {\bibfnamefont {Y.}~\bibnamefont {Zheng}}, \bibinfo {author}
  {\bibfnamefont {D.~F.}\ \bibnamefont {Xu}}, \bibinfo {author} {\bibfnamefont
  {X.~H.}\ \bibnamefont {Niu}}, \bibinfo {author} {\bibfnamefont {X.~P.}\
  \bibnamefont {Shen}}, \bibinfo {author} {\bibfnamefont {H.~F.}\ \bibnamefont
  {Yang}}, \bibinfo {author} {\bibfnamefont {P.}~\bibnamefont {Dudin}},
  \bibinfo {author} {\bibfnamefont {T.~K.}\ \bibnamefont {Kim}}, \bibinfo
  {author} {\bibfnamefont {M.}~\bibnamefont {Hoesch}}, \bibinfo {author}
  {\bibfnamefont {I.}~\bibnamefont {Vobornik}}, \bibinfo {author}
  {\bibfnamefont {Z.-A.}\ \bibnamefont {Xu}}, \bibinfo {author} {\bibfnamefont
  {X.~G.}\ \bibnamefont {Wan}}, \bibinfo {author} {\bibfnamefont {D.~L.}\
  \bibnamefont {Feng}},\ and\ \bibinfo {author} {\bibfnamefont {D.~W.}\
  \bibnamefont {Shen}},\ }\bibfield  {title} {\bibinfo {title} {Bulk and
  surface electronic structure of hexagonal structured {${\mathrm{PtBi}}_{2}$}
  studied by angle-resolved photoemission spectroscopy},\ }\href@noop {}
  {\bibfield  {journal} {\bibinfo  {journal} {Phys. Rev. B}\ }\textbf {\bibinfo
  {volume} {94}},\ \bibinfo {pages} {235140} (\bibinfo {year}
  {2016})}\BibitemShut {NoStop}%
\bibitem [{\citenamefont {Xu}\ \emph {et~al.}(2016)\citenamefont {Xu},
  \citenamefont {Xing}, \citenamefont {Xu}, \citenamefont {Li}, \citenamefont
  {Chen}, \citenamefont {Che}, \citenamefont {Lu}, \citenamefont {Dai},\ and\
  \citenamefont {Shi}}]{xu2016}%
  \BibitemOpen
  \bibfield  {author} {\bibinfo {author} {\bibfnamefont {C.~Q.}\ \bibnamefont
  {Xu}}, \bibinfo {author} {\bibfnamefont {X.~Z.}\ \bibnamefont {Xing}},
  \bibinfo {author} {\bibfnamefont {X.}~\bibnamefont {Xu}}, \bibinfo {author}
  {\bibfnamefont {B.}~\bibnamefont {Li}}, \bibinfo {author} {\bibfnamefont
  {B.}~\bibnamefont {Chen}}, \bibinfo {author} {\bibfnamefont {L.~Q.}\
  \bibnamefont {Che}}, \bibinfo {author} {\bibfnamefont {X.}~\bibnamefont
  {Lu}}, \bibinfo {author} {\bibfnamefont {J.}~\bibnamefont {Dai}},\ and\
  \bibinfo {author} {\bibfnamefont {Z.~X.}\ \bibnamefont {Shi}},\ }\bibfield
  {title} {\bibinfo {title} {Synthesis, physical properties, and band structure
  of the layered bismuthide {${\mathrm{PtBi}}_{2}$}},\ }\href@noop {}
  {\bibfield  {journal} {\bibinfo  {journal} {Phys. Rev. B}\ }\textbf {\bibinfo
  {volume} {94}},\ \bibinfo {pages} {165119} (\bibinfo {year}
  {2016})}\BibitemShut {NoStop}%
\bibitem [{\citenamefont {Kaiser}\ \emph {et~al.}(2014)\citenamefont {Kaiser},
  \citenamefont {Baranov},\ and\ \citenamefont {Ruck}}]{kaiser}%
  \BibitemOpen
  \bibfield  {author} {\bibinfo {author} {\bibfnamefont {M.}~\bibnamefont
  {Kaiser}}, \bibinfo {author} {\bibfnamefont {A.~I.}\ \bibnamefont
  {Baranov}},\ and\ \bibinfo {author} {\bibfnamefont {M.}~\bibnamefont
  {Ruck}},\ }\bibfield  {title} {\bibinfo {title} {{Bi$_2$Pt(hP9)} by
  low-temperature reduction of {Bi$_13$Pt$_3$I$_7$}: Reinvestigation of the
  crystal structure and chemical bonding analysis},\ }\href@noop {} {\bibfield
  {journal} {\bibinfo  {journal} {Zeitschrift für anorganische und allgemeine
  Chemie}\ }\textbf {\bibinfo {volume} {640}},\ \bibinfo {pages} {2742}
  (\bibinfo {year} {2014})}\BibitemShut {NoStop}%
\bibitem [{\citenamefont {Gao}\ \emph {et~al.}(2018)\citenamefont {Gao},
  \citenamefont {Zhu}, \citenamefont {Zheng}, \citenamefont {Wu}, \citenamefont
  {Zhang}, \citenamefont {Xi}, \citenamefont {Zhang}, \citenamefont {Zhang},
  \citenamefont {Hao}, \citenamefont {Ning},\ and\ \citenamefont
  {Tian}}]{Gao2018}%
  \BibitemOpen
  \bibfield  {author} {\bibinfo {author} {\bibfnamefont {W.}~\bibnamefont
  {Gao}}, \bibinfo {author} {\bibfnamefont {X.}~\bibnamefont {Zhu}}, \bibinfo
  {author} {\bibfnamefont {F.}~\bibnamefont {Zheng}}, \bibinfo {author}
  {\bibfnamefont {M.}~\bibnamefont {Wu}}, \bibinfo {author} {\bibfnamefont
  {J.}~\bibnamefont {Zhang}}, \bibinfo {author} {\bibfnamefont
  {C.}~\bibnamefont {Xi}}, \bibinfo {author} {\bibfnamefont {P.}~\bibnamefont
  {Zhang}}, \bibinfo {author} {\bibfnamefont {Y.}~\bibnamefont {Zhang}},
  \bibinfo {author} {\bibfnamefont {N.}~\bibnamefont {Hao}}, \bibinfo {author}
  {\bibfnamefont {W.}~\bibnamefont {Ning}},\ and\ \bibinfo {author}
  {\bibfnamefont {M.}~\bibnamefont {Tian}},\ }\bibfield  {title} {\bibinfo
  {title} {A possible candidate for triply degenerate point fermions in
  trigonal layered {PtBi}$_2$},\ }\href@noop {} {\bibfield  {journal} {\bibinfo
   {journal} {Nature Communications}\ }\textbf {\bibinfo {volume} {9}}
  (\bibinfo {year} {2018})}\BibitemShut {NoStop}%
\bibitem [{\citenamefont {Brese}\ and\ \citenamefont {von
  Schnering}(1994)}]{brese1994bonding}%
  \BibitemOpen
  \bibfield  {author} {\bibinfo {author} {\bibfnamefont {N.~E.}\ \bibnamefont
  {Brese}}\ and\ \bibinfo {author} {\bibfnamefont {H.~G.}\ \bibnamefont {von
  Schnering}},\ }\bibfield  {title} {\bibinfo {title} {Bonding trends in
  pyrites and a reinvestigation of the structures of {PdAs$_2$}, {PdSb$_2$},
  {PtSb$_2$} and {PtBi$_2$}},\ }\href@noop {} {\bibfield  {journal} {\bibinfo
  {journal} {Zeitschrift für anorganische und allgemeine Chemie}\ }\textbf
  {\bibinfo {volume} {620}},\ \bibinfo {pages} {393} (\bibinfo {year}
  {1994})}\BibitemShut {NoStop}%
\bibitem [{\citenamefont {Canfield}\ \emph {et~al.}(2016)\citenamefont
  {Canfield}, \citenamefont {Kong}, \citenamefont {Kaluarachchi},\ and\
  \citenamefont {Jo}}]{canfield}%
  \BibitemOpen
  \bibfield  {author} {\bibinfo {author} {\bibfnamefont {P.~C.}\ \bibnamefont
  {Canfield}}, \bibinfo {author} {\bibfnamefont {T.}~\bibnamefont {Kong}},
  \bibinfo {author} {\bibfnamefont {U.~S.}\ \bibnamefont {Kaluarachchi}},\ and\
  \bibinfo {author} {\bibfnamefont {N.~H.}\ \bibnamefont {Jo}},\ }\bibfield
  {title} {\bibinfo {title} {Use of frit-disc crucibles for routine and
  exploratory solution growth of single crystalline samples},\ }\href@noop {}
  {\bibfield  {journal} {\bibinfo  {journal} {Philosophical Magazine}\ }\textbf
  {\bibinfo {volume} {96}},\ \bibinfo {pages} {84} (\bibinfo {year}
  {2016})}\BibitemShut {NoStop}%
\bibitem [{\citenamefont {Rodríguez-Carvajal}(1993)}]{fullprof}%
  \BibitemOpen
  \bibfield  {author} {\bibinfo {author} {\bibfnamefont {J.}~\bibnamefont
  {Rodríguez-Carvajal}},\ }\bibfield  {title} {\bibinfo {title} {Recent
  advances in magnetic structure determination by neutron powder diffraction},\
  }\href@noop {} {\bibfield  {journal} {\bibinfo  {journal} {Physica B:
  Condensed Matter}\ }\textbf {\bibinfo {volume} {192}},\ \bibinfo {pages} {55
  } (\bibinfo {year} {1993})}\BibitemShut {NoStop}%
\bibitem [{\citenamefont {Pet{\v{r}}{\'{\i}}{\v{c}}ek}\ \emph
  {et~al.}(2014)\citenamefont {Pet{\v{r}}{\'{\i}}{\v{c}}ek}, \citenamefont
  {Du{\v{s}}ek},\ and\ \citenamefont {Palatinus}}]{jana}%
  \BibitemOpen
  \bibfield  {author} {\bibinfo {author} {\bibfnamefont {V.}~\bibnamefont
  {Pet{\v{r}}{\'{\i}}{\v{c}}ek}}, \bibinfo {author} {\bibfnamefont
  {M.}~\bibnamefont {Du{\v{s}}ek}},\ and\ \bibinfo {author} {\bibfnamefont
  {L.}~\bibnamefont {Palatinus}},\ }\bibfield  {title} {\bibinfo {title}
  {Crystallographic computing system {JANA}2006: General features},\
  }\href@noop {} {\bibfield  {journal} {\bibinfo  {journal} {Zeitschrift
  f\"{u}r Kristallographie - Crystalline Materials}\ }\textbf {\bibinfo
  {volume} {229}} (\bibinfo {year} {2014})}\BibitemShut {NoStop}%
\bibitem [{\citenamefont {Koepernik}\ and\ \citenamefont
  {Eschrig}(1999)}]{kopernik}%
  \BibitemOpen
  \bibfield  {author} {\bibinfo {author} {\bibfnamefont {K.}~\bibnamefont
  {Koepernik}}\ and\ \bibinfo {author} {\bibfnamefont {H.}~\bibnamefont
  {Eschrig}},\ }\bibfield  {title} {\bibinfo {title} {Full-potential
  nonorthogonal local-orbital minimum-basis band-structure scheme},\
  }\href@noop {} {\bibfield  {journal} {\bibinfo  {journal} {Physical review
  B}\ }\textbf {\bibinfo {volume} {59}} (\bibinfo {year} {1999})}\BibitemShut
  {NoStop}%
\bibitem [{\citenamefont {Perdew}\ \emph {et~al.}(1996)\citenamefont {Perdew},
  \citenamefont {Burke},\ and\ \citenamefont
  {Ernzerhof}}]{PhysRevLett.77.3865}%
  \BibitemOpen
  \bibfield  {author} {\bibinfo {author} {\bibfnamefont {J.~P.}\ \bibnamefont
  {Perdew}}, \bibinfo {author} {\bibfnamefont {K.}~\bibnamefont {Burke}},\ and\
  \bibinfo {author} {\bibfnamefont {M.}~\bibnamefont {Ernzerhof}},\ }\bibfield
  {title} {\bibinfo {title} {Generalized gradient approximation made simple},\
  }\href {https://doi.org/10.1103/PhysRevLett.77.3865} {\bibfield  {journal}
  {\bibinfo  {journal} {Phys. Rev. Lett.}\ }\textbf {\bibinfo {volume} {77}},\
  \bibinfo {pages} {3865} (\bibinfo {year} {1996})}\BibitemShut {NoStop}%
\bibitem [{\citenamefont {Cortijo}\ \emph {et~al.}(2016)\citenamefont
  {Cortijo}, \citenamefont {Kharzeev}, \citenamefont {Landsteiner},\ and\
  \citenamefont {Vozmediano}}]{PhysRevB.94.241405}%
  \BibitemOpen
  \bibfield  {author} {\bibinfo {author} {\bibfnamefont {A.}~\bibnamefont
  {Cortijo}}, \bibinfo {author} {\bibfnamefont {D.}~\bibnamefont {Kharzeev}},
  \bibinfo {author} {\bibfnamefont {K.}~\bibnamefont {Landsteiner}},\ and\
  \bibinfo {author} {\bibfnamefont {M.~A.~H.}\ \bibnamefont {Vozmediano}},\
  }\bibfield  {title} {\bibinfo {title} {Strain-induced chiral magnetic effect
  in weyl semimetals},\ }\href {https://doi.org/10.1103/PhysRevB.94.241405}
  {\bibfield  {journal} {\bibinfo  {journal} {Phys. Rev. B}\ }\textbf {\bibinfo
  {volume} {94}},\ \bibinfo {pages} {241405} (\bibinfo {year}
  {2016})}\BibitemShut {NoStop}%
\bibitem [{\citenamefont {Chen}\ \emph {et~al.}(2016)\citenamefont {Chen},
  \citenamefont {Luo}, \citenamefont {Xiao}, \citenamefont {Lu}, \citenamefont
  {Zhang}, \citenamefont {Yang}, \citenamefont {Li}, \citenamefont {Pei},
  \citenamefont {Shao}, \citenamefont {Zhang}, \citenamefont {Ling},
  \citenamefont {Xi}, \citenamefont {Song},\ and\ \citenamefont
  {Sun}}]{doi:10.1063/1.4947433}%
  \BibitemOpen
  \bibfield  {author} {\bibinfo {author} {\bibfnamefont {F.~C.}\ \bibnamefont
  {Chen}}, \bibinfo {author} {\bibfnamefont {X.}~\bibnamefont {Luo}}, \bibinfo
  {author} {\bibfnamefont {R.~C.}\ \bibnamefont {Xiao}}, \bibinfo {author}
  {\bibfnamefont {W.~J.}\ \bibnamefont {Lu}}, \bibinfo {author} {\bibfnamefont
  {B.}~\bibnamefont {Zhang}}, \bibinfo {author} {\bibfnamefont {H.~X.}\
  \bibnamefont {Yang}}, \bibinfo {author} {\bibfnamefont {J.~Q.}\ \bibnamefont
  {Li}}, \bibinfo {author} {\bibfnamefont {Q.~L.}\ \bibnamefont {Pei}},
  \bibinfo {author} {\bibfnamefont {D.~F.}\ \bibnamefont {Shao}}, \bibinfo
  {author} {\bibfnamefont {R.~R.}\ \bibnamefont {Zhang}}, \bibinfo {author}
  {\bibfnamefont {L.~S.}\ \bibnamefont {Ling}}, \bibinfo {author}
  {\bibfnamefont {C.~Y.}\ \bibnamefont {Xi}}, \bibinfo {author} {\bibfnamefont
  {W.~H.}\ \bibnamefont {Song}},\ and\ \bibinfo {author} {\bibfnamefont
  {Y.~P.}\ \bibnamefont {Sun}},\ }\bibfield  {title} {\bibinfo {title}
  {Superconductivity enhancement in the s-doped weyl semimetal candidate
  mote2},\ }\href {https://doi.org/10.1063/1.4947433} {\bibfield  {journal}
  {\bibinfo  {journal} {Applied Physics Letters}\ }\textbf {\bibinfo {volume}
  {108}},\ \bibinfo {pages} {162601} (\bibinfo {year} {2016})},\ \Eprint
  {https://arxiv.org/abs/https://doi.org/10.1063/1.4947433}
  {https://doi.org/10.1063/1.4947433} \BibitemShut {NoStop}%
\bibitem [{\citenamefont {Veyrat}\ \emph {et~al.}(2020)\citenamefont {Veyrat},
  \citenamefont {Labracherie}, \citenamefont {Bashlakov}, \citenamefont
  {Caglieris}, \citenamefont {Facio}, \citenamefont {Shipunov}, \citenamefont
  {Graf}, \citenamefont {Schoop}, \citenamefont {Giraud}, \citenamefont
  {van~den Brink}, \citenamefont {B\"uchner}, \citenamefont {Hess},
  \citenamefont {Aswartham},\ and\ \citenamefont {Dufouleur}}]{inprep}%
  \BibitemOpen
  \bibfield  {author} {\bibinfo {author} {\bibfnamefont {A.}~\bibnamefont
  {Veyrat}}, \bibinfo {author} {\bibfnamefont {V.}~\bibnamefont {Labracherie}},
  \bibinfo {author} {\bibfnamefont {D.~L.}\ \bibnamefont {Bashlakov}}, \bibinfo
  {author} {\bibfnamefont {F.}~\bibnamefont {Caglieris}}, \bibinfo {author}
  {\bibfnamefont {J.~I.}\ \bibnamefont {Facio}}, \bibinfo {author}
  {\bibfnamefont {G.}~\bibnamefont {Shipunov}}, \bibinfo {author}
  {\bibfnamefont {L.}~\bibnamefont {Graf}}, \bibinfo {author} {\bibfnamefont
  {J.}~\bibnamefont {Schoop}}, \bibinfo {author} {\bibfnamefont
  {R.}~\bibnamefont {Giraud}}, \bibinfo {author} {\bibfnamefont
  {J.}~\bibnamefont {van~den Brink}}, \bibinfo {author} {\bibfnamefont
  {B.}~\bibnamefont {B\"uchner}}, \bibinfo {author} {\bibfnamefont
  {C.}~\bibnamefont {Hess}}, \bibinfo {author} {\bibfnamefont {S.}~\bibnamefont
  {Aswartham}},\ and\ \bibinfo {author} {\bibfnamefont {J.}~\bibnamefont
  {Dufouleur}},\ }\bibfield  {title} {\bibinfo {title} {Topology and
  two-dimesional superconductivity in trigonal {PtBi$_2$}}} (\bibinfo {year}
  {2020}),\ \bibinfo {note} {in preparation}\BibitemShut {NoStop}%
\end{thebibliography}%

\newpage{}

\newpage{}

\end{document}